\documentclass[fleqn,usenatbib]{mnras}

\usepackage{newtxtext,newtxmath}

\usepackage[T1]{fontenc}
\usepackage{ae,aecompl}

\usepackage{graphicx}	
\usepackage{amsmath}	
\usepackage{amssymb}	

\title[Unique compact lensing cluster CLIO]{MUSE spectroscopy and deep observations of a unique compact JWST target, lensing cluster CLIO}

\author[A. Griffiths~et~al.~]
{Alex~Griffiths,$^{1}$\thanks{E-mail: \href{mailto:alex.griffiths@nottingham.ac.uk}{alex.griffiths@nottingham.ac.uk}}
Christopher~J.~Conselice,$^{1}$
Mehmet~Alpaslan,$^{2}$
Brenda~L.~Frye,$^{3}$ 
\newauthor
Jose~M.~Diego,$^{4}$
Adi~Zitrin,$^{5}$
Haojing~Yan,$^{6}$
Zhiyuan~Ma,$^{6}$
Robert Barone-Nugent,$^{7}$
\newauthor
Rachana Bhatawdekar,$^{1}$
Simon.~P.~Driver,$^{8,9}$
Aaron~S.~G.~Robotham,$^{9}$ 
\newauthor
Rogier~A.~Windhorst$^{10}$ and J.~Stuart~B.~Wyithe$^{7}$
\\
$^{1}$School of Physics and Astronomy, The University of Nottingham, University Park, Nottingham NG7 2RD, UK\\
$^{2}$Center for Cosmology and Particle Physics, 
New York University, 726 Broadway, New York, NY 10003, USA\\
$^{3}$Department of Astronomy/Steward Observatory, University of Arizona, 933 North Cherry Avenue, Tucson, AZ 85721, USA\\
$^{4}$IFCA, Instituto de F\'isica de Cantabria (UC-CSIC), Av. de Los Castros s/n, 39005 Santander, Spain\\
$^{5}$Physics Department, Ben-Gurion University of the Negev, PO Box 653, Beer-Sheva 84105, Israel\\
$^{6}$Department of Physics and Astronomy, University of Missouri, Columbia, MO 65211, USA\\
$^{7}$School of Physics, University of Melbourne, Parkville, Victoria, VIC 3010, Australia\\
$^{8}$SUPA, School of Physics and Astronomy, University of St Andrews, North Haugh, St Andrews, KY16 9SS, UK\\
$^{9}$ICRAR, The University of Western Australia, 35 Stirling Highway, Crawley, WA 6009, Australia\\
$^{10}$School of Earth and Space Exploration, Arizona State University, Tempe, AZ 85287-1404, USA
}

\date{Accepted XXX. Received YYY; in original form ZZZ}

\pubyear{2017}

\begin{document}
\label{firstpage}
\pagerange{\pageref{firstpage}--\pageref{lastpage}}
\maketitle

\begin{abstract}
We present the results of a VLT MUSE/FORS2 and Spitzer survey of a unique compact lensing cluster CLIO at z = 0.42, discovered through the GAMA survey using spectroscopic redshifts. Compact and massive clusters such as this are understudied, but provide a unique prospective on dark matter distributions and for finding background lensed high-z galaxies.  The CLIO cluster was identified for follow up observations due to its almost unique combination of high mass and dark matter halo concentration, as well as having observed lensing arcs from ground based images. Using dual band optical and infra-red imaging from FORS2 and Spitzer, in combination with MUSE optical spectroscopy we identify 89 cluster members and find background sources out to z = 6.49. We describe the physical state of this cluster, finding a strong correlation between environment and galaxy spectral type. Under the assumption of a NFW profile, we measure the total mass of CLIO to be M$_{200} = (4.49 \pm 0.25) \times 10^{14}$ M$_\odot$. We build and present an initial strong-lensing model for this cluster, and measure a relatively low intracluster light (ICL) fraction of 7.21 $\pm$ 1.53\% through galaxy profile fitting. Due to its strong potential for lensing background galaxies and its low ICL, the CLIO cluster will be a target for our 110 hour JWST `Webb Medium-Deep Field' (WMDF) GTO program.
\end{abstract}




\section{Introduction}

Formed via the gravitation collapse of overdensities in the initial primordial density field, galaxy clusters are some of the largest gravitationally bound objects in the Universe. Clusters can serve as unique laboratories for the study of both astrophysics, and cosmology. Since they were first hinted at in the clustering of nebula seen by the Herschels, galaxy clusters have led to a number of groundbreaking scientific discoveries. For example, in 1933, Zwicky demonstrated the need for a dark matter component by investigating the mass of the Coma cluster \citep{Zwicky1933}. In terms of galaxy evolution, in the early 1950s Spitzer and Baade first studied cluster environments to reveal collisional stripping \citep{Spitzer1951}.

In recent decades, advances in observational capabilities have provided a detailed insight into the structure of galaxy clusters and their dynamical evolution. This improved understanding of cluster physics has provided a tool to investigate the large-scale structure, refine cosmological models, and probe the early Universe through gravitational lensing \citep[e.g.,][]{Zheng2012,Coe2015,Dye2015}. With the advent of new observational facilities such as JWST, Euclid and LSST in the next decade, clusters are bound to become an even more important environment to learn new astrophysics. JWST observations of lensing clusters will probe significantly deeper than current observational capabilities in the search for high redshift galaxies and population III stars. While Euclid and LSST will use galaxy clusters to compliment studies into the accelerated expansion of the universe and the nature of dark energy and dark matter.

A prime example of this is through the exploitation of massive galaxy clusters as gravitational lenses. This has proved to be a valuable method for studying the early Universe. Typical methods such as the Lyman break technique \citep[e.g.,][]{Steidel1996,Giavalisco2004} often rely on the use of deep blank fields, in which early galaxy populations have been characterised through UV colours, stellar masses and ages \citep{Duncan2014}. The magnification of faint background sources by gravitational lenses provides a complementary method to detect even fainter galaxies over a wide range of redshifts \mbox{\citep[e.g.,][]{Brammer2012,Alavi2014}}. In this vein, large, dedicated cluster surveys such as the Hubble Frontier Fields \cite[HFF;][]{Lotz2014,Koekemoer2014} and the Cluster Lensing And Supernova survey with Hubble \citep[CLASH;][]{Postman2012} have targeted a number of massive clusters for use in gravitational lensing studies. 
  
Detailed observations and programs such as the HFF and CLASH have led to significant improvements in our ability to successfully model cluster mass distributions. Through the identification of lensed arcs, and multiply-imaged background sources, it is possible to refine mass models such that clusters can be used as gravitational telescopes to locate, and study high redshift objects. These density maps are constructed as a combination of the cluster population, and an inferred dark matter distribution \citep[e.g.,][]{Frye1998,Zitrin2012b,Jauzac2015,Grillo2015,Kawamata2016}. 

By far the most common type of cluster in the distant universe that has been studied in detail are the most massive ones, such as those in the Frontier Fields and CLASH. There are however good reasons for wanting to investigate other types of clusters. In particular, compact but massive clusters provide a potentially powerful approach for obtaining new and complementary information about dark matter distributions, as well as an alternative approach for studying gravitationally lensed galaxies. The reasons for this are that a compact cluster will have a potentially different structure than a typical large massive cluster, providing a new avenue to study dark matter distributions. Furthermore, a compact massive cluster will lens background galaxies through a lower amount of intracluster light, making them ideal targets for finding the most distant galaxies in the universe.

Here, we present the first results obtained from dedicated observations of such a compact galaxy cluster, which we name the CLIO cluster after the Greek muse of history (see Section~\ref{sec:target} for more details). CLIO is identified within the Galaxy and Mass Assembly \cite[GAMA;][]{Driver2011,Liske2015} survey at an intermediate redshift of $z = 0.42$. We use dual band FORS2 and IRAC imaging in conjunction with spectroscopic MUSE data in order to refine the properties of this cluster, building upon the initial measurements from the GAMA data. We measure spectroscopic redshifts to identify cluster members, and aim to identify multiply-imaged sources in order to construct an improved mass distribution of the cluster. Combining spectroscopy with imaging data also enables us to measure member luminosities and masses in order to estimate the relative baryon mass fraction of the cluster.

As mentioned, high mass clusters which provide sufficient curvature in spacetime to be beneficial for gravitational lensing, also correlate with high rates of diffuse intracluster light (ICL), which can prove to be a hindrance, as while this component can be subtracted off it leaves high levels of residual noise. Therefore minimal ICL contamination is essential when studying cluster lenses where such straylight components can impede the detection of faint lensed sources in the cluster centre, which would otherwise provide crucial constraints on mass models. Recent studies have found ICL fractions at medium redshifts (0.3 < z < 0.6) of $\sim$10-20\%, around half of the local value \citep[$\sim$ 40\%;][]{Gonzalez2013}, suggesting that the ICL is still being produced at z < 1 \citep{Morishita2016,Jimenez2016}. Thus, in this study we investigate the ICL fraction in regards to utilising the CLIO cluster as a potential gravitational lens.

This study is part of preliminary work towards our James Webb Space Telescope (JWST) GTO program 1176 (PI Windhorst), `Webb Medium-Deep Field' survey (WMDF). This program will look at a combination of lensing clusters and blank fields in the search for first light objects. While there are a number of well studied lensing clusters, they are not all ideal candidates for the search of first light objects due to strong ICL contamination or high Zodiacal backgrounds. We will thus, compliment the existing HFF and CLASH lensing samples with recently discovered, high mass, compact clusters such as the CLIO cluster.

The layout of this paper is as follows. In Section~\ref{sec:target} we discuss the group finding algorithm, including how CLIO was identified as an unique viable target for study. Section~\ref{sec:obs} gives an overview of the observations, including data reduction and spectral analysis. In Section~\ref{sec:res} we derive key physical properties of the CLIO cluster and its member galaxies and perform preliminary lensing, and ICL analysis. Finally, we summarise our main results and draw conclusions in Section~\ref{sec:summary}. Throughout this paper we adopt a $\Lambda$ cold dark matter cosmological model with $\Omega_{\Lambda} = 0.7$, $\Omega_{\textrm{M}} = 0.3$ and $H = 70$ km s$^{-1}$ Mpc$^{-1}$. At the cluster redshift of z = 0.42, 1\arcsec~corresponds to a physical distance of 5.533 kpc. All magnitudes are given in the AB system \citep{Oke1974}.


\section{Target Selection}
\label{sec:target}

GAMA is a deep spectroscopic survey over $\sim$250 square degrees down to r $< 19.8$ mag, providing the largest, highly complete sample of low mass galaxy groups in the local Universe. CLIO is selected from v8 of the GAMA Galaxy Group Catalogue \citep[G$^3$C;][]{Robotham2011}. Groups in the G$^3$C are identified using a modified friends-of-friends (FoF) algorithm, which is run on the same distribution of galaxies projected onto the sky, as well as along the line of sight. Galaxies identified as linked to each other in both projections are considered to be grouped together; this successfully eliminates any redshift space distortion effects. Parameters for the FoF algorithm are calibrated and optimised by running it on a series of bespoke mock galaxy catalogues whose geometry and luminosity function match those of the GAMA survey \citep{Merson2013}.

\begin{figure}
	\includegraphics[width=\columnwidth]{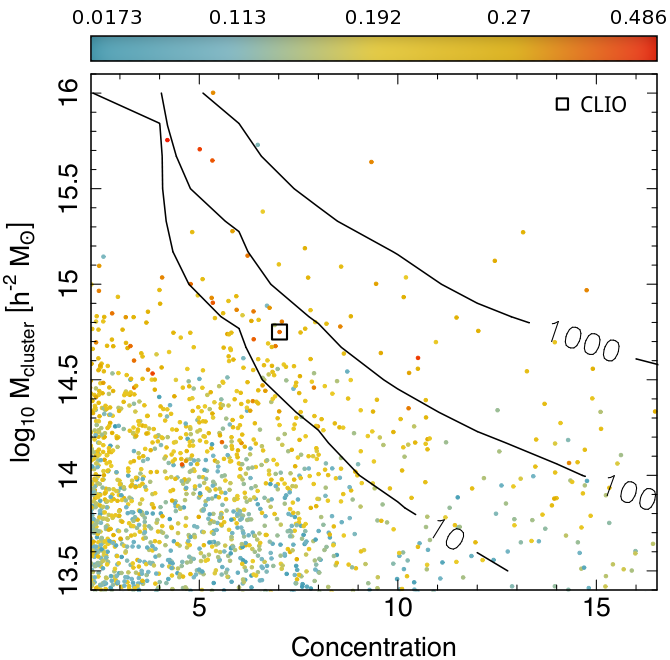}
	\caption{Dynamical masses as a function of concentration for GAMA groups, with each point coloured by the group's median redshift. Contours represent the estimated number of lensed galaxies between $1 \leq z \leq 8$. CLIO is identified by the black square.}
	\label{fig:concDat}
\end{figure}

We analyse galaxy groups/clusters from the G$^3$C based on a combination of various selection criteria. Selection is initially based on cluster concentration, since it may be easier to remove JWST straylight gradients in compact clusters with lower ICL fractions. Concentration estimates are derived by fitting a Navarro-Frenk-White \citep[NFW;][]{Navarro1996} profile to the enclosed mass within the 50$^{\rm{th}}$, and 68$^{\rm{th}}$ percentile radii, with concentration as a free parameter. We model the number of expected lensed sources between $1 \leq z \leq 8$ as a function of group mass and concentrations, calculated assuming a singular isothermal sphere (SIS) density profile, with magnification bias and high redshift galaxy densities based on the luminosity functions of \citet{Bouwens2015}. The lensed image rates are shown as overlaid contours in Figure~\ref{fig:concDat}. 

We require a cluster redshift that is optimal for lensing studies of high redshift galaxies with JWST. We find that the cluster redshift must be kept to $z \lesssim 0.5$ in order to minimize the effects of the cluster spectral energy distribution (SED) in the wavelength range of $\lambda = 3-5\mu m$, where the JWST Near-Infrared Camera (NIRCam) is the most sensitive and the Zodiacal background is darkest. The increased ICL surface brightness at these lower redshifts is compensated for by lower, near-IR K-corrections of similar clusters at higher redshifts.

Clusters are further sub-selected from the sample based on the caustic mass estimation technique \citep{Diaferio1999,Alpaslan2012}. Cluster members are analysed in a redshift-space diagram in which the projected radial separation from the group centre is plotted against the line-of-sight velocity relative to the group centre of mass. Within this phase-space, the spherical infall model \citep{Regos1989} predicts the existence or trumpet shaped caustics which define the escape velocity of the cluster as a function of radial distance from the cluster centre. Thus, galaxies which fall outside of the caustics are not considered to be gravitationally bound to the cluster. Through this analysis we select clusters with optimal distribution within this phase-space. Finally, we examine the redshift distribution of cluster members to check for potential merging systems or line-of-sight alignments, which complicate lensing studies.

Original identification of the CLIO cluster in version 5 of the G$^3$C is GAMA100033. However, more recent versions (v6 and onwards) label the cluster as GAMA100050. To avoid any ID ambiguity, we provide a unique name for the cluster. We take our inspiration from the Multi Unit Spectroscopic Explorer (MUSE) on which a vast majority of this preliminary work is based. We name the cluster after one of the nine muses of Greek mythology, the muse of history, Clio.    

We identified this cluster as a promising candidate for follow-up observations; based on its unique combination of implied high mass and concentration, as well as the detected presence of lensing arcs in ground based imaging. CLIO has a median redshift of z = 0.42 and is estimated to have a dynamical mass of $\sim 6 \times 10^{14}$ M$_\odot$, and a concentration of 6.3. From GAMA there are 14 spectroscopically confirmed members and the central galaxy ID is 323174 (G$^3$Cv9). The cluster has an initial velocity dispersion of 633 km s$^{-1}$ and a radius of R$_{50}$ = 0.66 Mpc. This clusters combination of high mass and concentration (which is sufficient to generate well resolved lenses, as evidenced by HST imaging), as well as a redshift that is favourable to lensing high redshift galaxies makes CLIO ideal for our targeted study. 


\section{Observations and Data Reduction}
\label{sec:obs}

Optical imaging and spectroscopy of the galaxy cluster CLIO were obtained as part of program 096.B-0605(A) (PI Conselice) at the European Southern Observatory (ESO) Paranal facility. Ancillary infrared observations were made as part of \textit{Spitzer} Cycle 13 Program 13024 (PI Yan). These observations were made as part of a study into the most dark matter dense galaxy clusters identified within the GAMA survey in an effort to study the properties of these unique objects.

\subsection{Optical Imaging}
Dual band photometric data were obtained with the FOcal Reducer/low dispersion Spectrograph (FORS2) multi mode instrument mounted on the VLT at ESO's Paranal Observatory \citep{Appenzeller1998}. CLIO was observed at a pointing of $\alpha =$ 08:42:21.2, $\delta =$ +01:38:26.3 using the FORS2 High Resolution Collimator equipped with a mosaic of two 2k x 4k CCDs, providing a pixel scale of 0.125 arcsec pixel$^{-1}$ over a 4.25$^{\prime}$ x 4.25$^{\prime}$ field of view (FoV). Observations were carried out on the nights of December 13th 2015 (12x580s) and January 15th 2016 (10x600s) in the R\_SPECIAL and g\_HIGH filters respectively. An earlier observation run with the g\_HIGH filter was initiated on the night of December 22nd 2015, however it was subsequently aborted due to sky saturation by the moon.    

The photometric data were reduced using the dedicated FORS2 instrument pipeline (v. 5.3.5), implemented within the ESO data reduction environment Reflex v. 2.8.5 \citep{Freudling2013}. The pipeline performs all standard reduction procedures including bias subtraction, cosmic rays removal, flat-field corrections and astrometry. Final science images are 2 x 2 binned, resulting in a final pixel scale of 0.25 arcsec pixel$^{-1}$ and a projected field of view of 8.3$^{\prime}$ x 8.3$^{\prime}$. However, due to vignetting the effective field of view is reduced to $\sim$ 6.8$^{\prime}$ x 6.8$^{\prime}$. Reduced science images are then background subtracted and median combined with WCS offsets using standard \textsc{iraf} tasks, imarith and imcombine.  

We performed aperture photometry using SE\textsc{xtractor} \citep{Bertin1996} in dual-image mode. We apply a 3 x 3 pixel Gaussian filter over both images and set a detection threshold of 0.5$\sigma$ for optimal identification of faint background sources. The FORS2 r-band image is used as a reference in which centroids and apertures are calculated for photometric measurements of both r, and g-band images. We use Kron-like elliptical apertures from SE\textsc{xtractor} in order to calculate final object magnitudes. Due to technical issues, standards were not taken for the g\_HIGH observations, so photometric calibrations of both bands are undertaken. We adjusted magnitudes to the Sloan Digital Sky Survey (SDSS) system using a set of standard stars within our observed field. We corrected instrumental magnitudes for atmospheric extinction\footnote{obtained from the ESO quality control database \url{http://www.eso.org/observing/dfo/quality/index_fors2.html}} and calculated photometric zeropoints. Finally, we calculate colour transformation coefficients in order to account for further filter offsets. 

We obtain additional data from the first public data release of the Hyper Suprime-Cam Subaru Strategic Program \citep[HSC-SSP;][]{Aihara2017b}. Along with g,r,i,z,y imaging we obtain catalogue data, including forced photometry and photometric redshift estimates. We utilise the g and r-band HSC photometry to perform additional checks on our photometric calibration and colour transformations.  

\subsection{IR Imaging}
We also obtained \textit{Spitzer} InfraRed Camera \citep[IRAC;][]{Fazio2004} observations in its 3.6 and 4.5~$\mu m$ channels. These data were taken on 2nd February 2016 as part of Cycle 13 Program ID 13024 (PI Yan). The observations were done in two Astronomical Data Requests (AORs), one for each channel. The two AORs were executed back-to-back, and hence the images in both channels essentially have the same orientation. The $5.2^{\prime}\times 5.2^{\prime}$ IRAC field-of-view is large enough to cover the entire cluster, and hence we only dithered around one pointing. IRAC takes images in both channels simultaneously: if one chooses to centre the target field in one channel, an adjacent flanking field will be automatically obtained at the same time in the other channel, and there is a separation of $1.52^{\prime}$ in between the edges of the two fields. We used 100-second frame time and medium-scale cycling dither of 120 positions, which resulted in a total integration of 12,000 seconds in both channels. The dithering scale is large enough such that the two field-of-views have no gaps after mosaicking. However, as the flanking fields in the two channels are at the opposite sides of the target field, only the target field itself has data in both channels.

\textit{Spitzer}{} Science Center (SSC) reduces all raw data for users through its standard pipeline to the point of producing Basic Calibrated Data (BCD), which are individual exposures removed of various instrumental effects (such as dark current subtraction, flat-fielding, etc.) In recent years, the pipeline has extended its capability to mitigate several artifacts commonly found in IRAC data (such as ``column pulldown'', ``banding'', etc.), and produces ``Corrected BCD'' (CBCD) images. Our further reduction of the data involved stacking the individual exposures, which was based on the CBCD images retrieved from the Spitzer Heritage Archive (SHA). For this purpose, we utilized the MOPEX software (v. 18.5.0) developed at SSC, which is capable of fine-tuning the astrometry of individual images, doing proper sub-sampling of the image and projection, and summing the images using a variety of algorithms. We removed the background of each CBCD image by subtracting its background map derived from the image using a median filter of $45\times 45$ (native) pixels in size. We adopted a pixel scale of 0.6\arcsec~(i.e., about half of the native pixel scale) for the final mosaics. The projection centre and the image dimensions were chosen such that the final mosaics in both channels would be precisely aligned. The stacking was done by a linear weighted sum of input pixels with weights equal to the area overlap with the output pixel, which is the default stacking scheme of MOPEX. In the process, 3$\sigma$ outliers were rejected before combining. 

\subsection{MUSE Optical Spectroscopy}
On the nights of the 12th (7x900s) and 13th December 2015 (7x900s), spectroscopic data of the CLIO cluster were obtained with the MUSE Integral Field Spectrograph located at the VLT \citep{MUSE}. The spectrograph offers a wavelength coverage of 480 - 930 nm with a mean spectral resolution of R $\sim$ 3000. The $\sim$1 arcmin$^2$ field of view is provided by MUSE's wide field mode, resulting in a spatial resolution of 0.2 arcsec pixel$^{-1}$.

The MUSE data were reduced and combined into a single datacube using the instrument pipeline (v. 1.6.2) in the ESO Reflex environment. The pipeline works on individual exposures, performing bias subtraction, flatfielding and wavelength calibration across the full wavelength range. Flux calibration is then performed using reference stars and the 14 individual exposures are combined into a single datacube with a resulting $\sim$1.8 arcmin$^2$ field of view. The datacube was further processed with the \textsc{zap} pipeline which makes use of the Zurich Atmosphere Purge \citep{Soto2016} code to remove residual sky contamination using principal component analysis.

\subsection{Spectral Analysis}
\label{sec:spec}
Initial redshift analysis was performed using the MUSE Line Emission Tracker\footnote{\textsc{muselet} is an open-source python package developed by the MUSE Consortium as part of the MUSE Python Data Analysis Framework (\textsc{mpdaf}) \url{http://mpdaf.readthedocs.io/en/latest/index.html}} (\textsc{muselet}). This tool splits the datacube into line-weighted pseudo-narrow band images of 6.25 \AA\ width across the full wavelength range of the MUSE datacube. The continuum is estimated from spectral medians of $\sim$25 \AA\ on either side of the narrow band region. SE\textsc{xtractor} is then used to detect line emission in the pseudo narrow-band images, a composite catalogue is created in which continuum sources can be isolated, and redshifts estimated. 

Object detection was performed by running SE\textsc{xtractor} on the FORS2 r-band image. We use the \textsc{iraf} task wregister, to transform the output segmentation map and the FORS2 r-band image to the same pixel scale and FoV as the MUSE datacube. The resulting spatial profiles provided by the segmentation map were then used as a basis to extract 1D weighted spectra and variance from the MUSE datacube using a custom \textsc{matlab} code. The chosen SE\textsc{xtractor} parameters caused some deblending of the most extended sources into multiple objects. Where this occurred, the objects were manually inspected and segmentations regions merged before extraction of spectra. Redshifts were then determined using a customised version of the Manual and Automatic Redshifting Software (\textsc{marz}), which uses a cross-correlation algorithm to match input spectra against a number of emission and absorption-line template spectra \citep{Hinton2016}. Customisation of the software was performed in order to include a further ten high redshift spectral templates. These additional templates were obtained from zCosmos \citep{Lilly2009}, the VIMOS-VLT Deep Survey \citep[VVDS;][]{LeFevre2013} and UDSz \citep[ESO Large Programme 180.A-0776, PI: Almaini,][private communication]{Bradshaw2013,McLure2013}. All reduced spectra are passed through the \textsc{marz} software and along with an initial redshift estimate, each spectrum is assigned a quality operator ($QOP$) based on the cross-correlation strength. Each spectrum is then visually inspected in order to improve redshift determinations where necessary and assign quality flags. Each redshift is assigned a final quality flag ($Q$) with values ranging from 1 to 5 based on the following specifications:
\begin{description}
\item 1: Unknown redshift with no evident spectral features, determined by cross-correlation only. 
\item 2: Possible but uncertain redshift, single undetermined spectral feature or low cross-correlation strength. 
\item 3: Probable redshifts with a single strong spectral feature or numerous faint absorption features.
\item 4: Secure redshifts with multiple strong spectral features present and a high cross-correlation strength.
\item 5: Secure redshift of non-extragalactic source matched to a stellar template. 
\end{description}

\begin{figure*}
	\includegraphics[width=\textwidth]{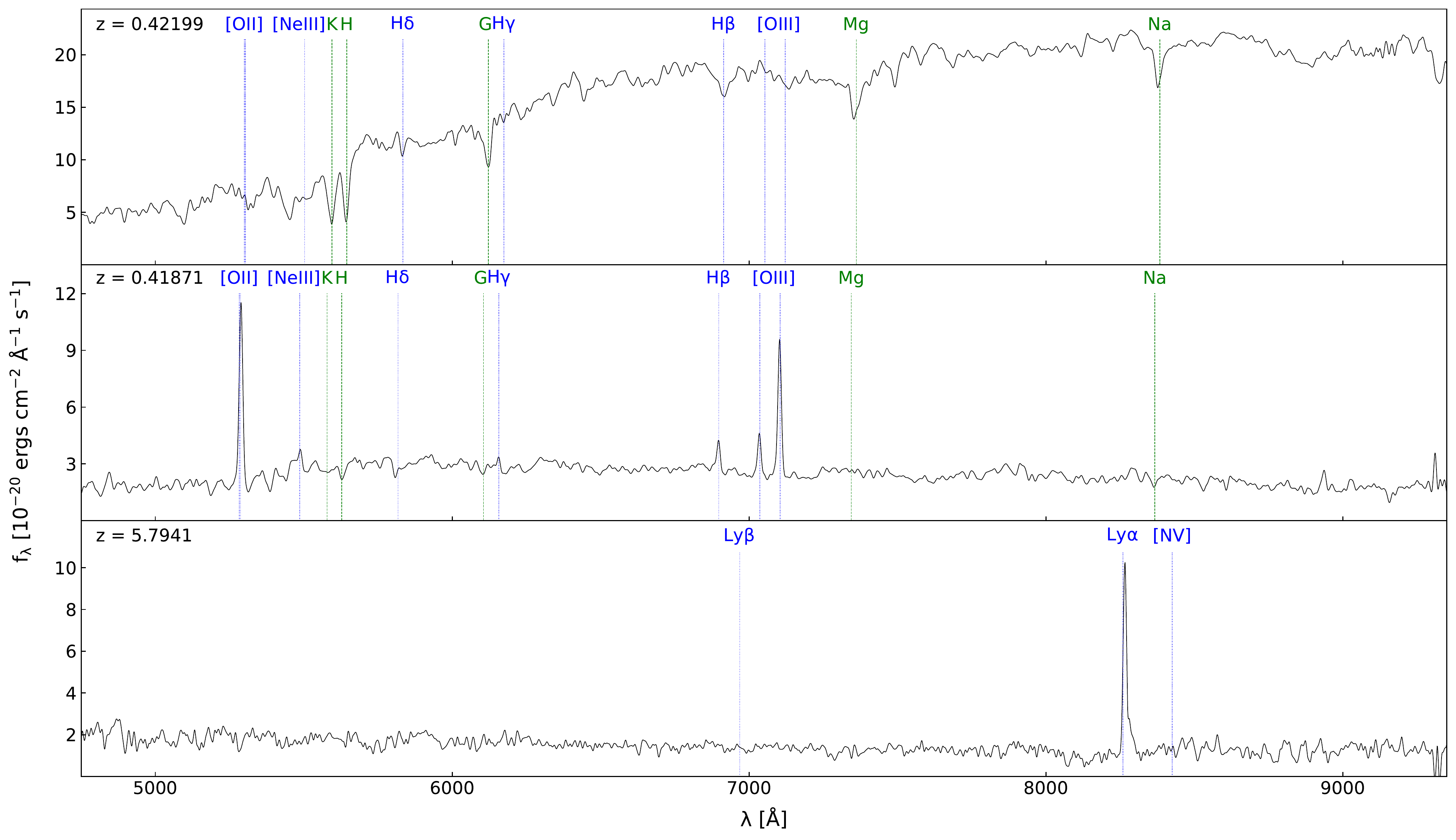}
    \caption{Examples of 1D spectra extracted from the MUSE datacube. From top to bottom, we show: a quality level 4 spectrum identified through multiple absorption features (marked by green dashed lines), a quality 4 cluster member spectrum based on multiple emission lines (mainly [\ion{O}{ii}] and [\ion{O}{iii}], marked by the blue dotted lines), and a quality 3 high redshift spectrum identified through the strong asymmetrical \ion{Ly}{$\alpha$} feature.}  
    \label{fig:spectra}
\end{figure*}

Emission line redshifts determined by \textsc{muselet} provide sufficient reference in order to manually fit a number of uncertain spectra within the \textsc{marz} GUI. Of the 44 \textsc{muselet} detections, 35 fit well to the initial \textsc{marz} estimates, 6 were used to manually adjust redshifts, and 3 were found to be spurious emission line detections. Additionally, 61 single emission line sources were identified by \textsc{muselet} which were used to confirm or reject fits to Ly$\alpha$ or [\ion{O}{iii}] lines. Typical examples of spectra extracted from the MUSE datacube can be seen in Figure~\ref{fig:spectra}.

A total of 406 objects were detected within our observed MUSE field, of these, 184 of the extracted spectra were of high enough quality for redshifts to be determined. A summary of our redshift detections can be found in Table~\ref{tab:spectra}. We construct individual redshift catalogues of cluster members and background objects with $Q = 3$ and $Q = 4$ spectra only (with an exception of lensing arcs discussed in Section~\ref{sec:lensmass}). Analysis described within this paper makes use of the results contained within this catalogue, a sample of catalogue entries and a description of columns is provided in Table~\ref{tab:clustercat} \& \ref{tab:bgcat}. 

\begin{table}
  \caption{Summary of redshift determinations and quality flags assigned to spectra extracted from the MUSE datacube.}
  \label{tab:spectra}
  \begin{tabular}{lccc}
    \hline
    $Q$ & Number & Fraction & Redshift Range \\
    \hline
    1 & 168 & 41.4\% & -\\
    2 & 54 & 14.0\% & -\\
    3 & 97 & 23.2\% & $0.0001 \leq z \leq 6.49$\\
    4 & 80 & 19.7\% & $0.08 \leq z \leq 1.22$\\
    5 & 7 & 1.7\% & -\\
    \hline
  \end{tabular}
\end{table}

\begin{table*}
  \caption{Cluster member spectroscopic redshift catalogue ordered by ID. This table has been truncated and is available in its entirety in the online version of this paper. ID's quoted are from this study and correspond to spectra extracted from the MUSE datacube. Columns 4 to 11 provide g,r,3.6$\mu m$ and 4.5$\mu m$ magnitudes derived in this study along with their corresponding errors. g,r,i,z and y magnitudes and their corresponding errors taken from the HSC catalogue \citep{Aihara2017a} are included in columns 12 to 21. We provide k-corrections (calculated following methods outlined in \citet{Chilingarian2010}) for g and r-band photometry in columns 22 and 23. Columns 24 and 25 provide stellar mass estimates detailed in Section~\ref{sec:gpop}, while columns 26 and 27 give DEmP photometric redshifts and their confidence values obtained from the HSC first data release \citep{Tanaka2017}. Finally, we provide spectroscopic redshifts and their quality flags as detailed in this section.}
  \label{tab:clustercat}
  \begin{tabular}{ccccccccccc}
    \hline
    id & ra & dec & mag\_g,... & hscmag\_g,... & kcorr\_g,... & mass\_g,... & photo\_z & photo\_zconf & spec\_z & Q\\
    & (deg) & (deg) & (mag) & (mag) & (mag) & (M$_{\odot}$) & & & &\\
    \hline
    23 & 130.596 & 1.654 & 22.99 $\pm$ 0.16 & 21.68 $\pm$ 0.01 & 1.47 & 1.57E+10 & 0.40 & 0.71 & 0.42 & 4\\
	28 & 130.591 & 1.654 & 23.16 $\pm$ 0.12 & 23.27 $\pm$ 0.02 & 1.47 & 1.32E+10 & 0.45 & 0.88 & 0.43 & 4\\
	34 & 130.589 & 1.653 & 22.67 $\pm$ 0.07 & 23.19 $\pm$ 0.01 & 1.55 & 2.61E+10 & 0.40 & 0.75 & 0.43 & 4\\
	36 & 130.588 & 1.653 & 24.31 $\pm$ 0.27 & 24.67 $\pm$ 0.04 & 1.47 & 4.69E+09 & 0.42 & 0.69 & 0.42 & 3\\
	41 & 130.584 & 1.653 & 24.21 $\pm$ 0.30 & 24.13 $\pm$ 0.03 & 1.52 & 5.95E+09 & 0.47 & 0.69 & 0.42 & 4\\
	42 & 130.595 & 1.653 & 21.87 $\pm$ 0.05 & 21.49 $\pm$ 0.00 & 1.60 & 6.95E+10 & 0.42 & 0.98 & 0.42 & 4\\
	43 & 130.580 & 1.653 & 22.98 $\pm$ 0.12 & 22.00 $\pm$ 0.01 & 1.42 & 1.39E+10 & 0.37 & 0.78 & 0.42 & 4\\
	45 & 130.578 & 1.652 & 22.77 $\pm$ 0.08 & 23.31 $\pm$ 0.01 & 1.45 & 1.83E+10 & 0.43 & 0.82 & 0.42 & 4\\
    \hline
  \end{tabular}
\end{table*}

\begin{table*}
  \caption{Background object spectroscopic redshift catalogue order by ID. This table has been truncated and is available in its entirety in the online version of this paper. Columns in this catalogue are identical to that of the cluster member catalogue with the exception of k-correction and mass estimations which are not included due to the redshift limitations of the methods used.}
  \label{tab:bgcat}
  \begin{tabular}{ccccccccc}
    \hline
    id & ra & dec & mag\_g,... & hscmag\_g,... & photo\_z & photo\_zconf & spec\_z & Q\\
    & (deg) & (deg) & (mag) & (mag) & & & &\\
    \hline
    5  & 130.591 & 1.627 & 25.52 $\pm$ 0.62 & 25.58 $\pm$ 0.08 & 0.36 & 0.81 & 6.09 & 3\\
    8  & 130.580 & 1.656 & 24.70 $\pm$ 0.30 & 25.38 $\pm$ 0.08 & 1.21 & 0.18 & 1.26 & 3\\
	17 & 130.588 & 1.655 & 22.98 $\pm$ 0.09 & 23.62 $\pm$ 0.02 & 0.58 & 0.90 & 0.55 & 4\\
	22 & 130.588 & 1.654 & 25.77 $\pm$ 0.77 & 26.11 $\pm$ 0.14 & 0.77 & 0.43 & 0.72 & 3\\
	25 & 130.586 & 1.654 & 26.51 $\pm$ 1.24 & 25.26 $\pm$ 0.09 & 0.97 & 0.43 & 0.97 & 3\\
	27 & 130.584 & 1.654 & 24.06 $\pm$ 0.20 & 24.66 $\pm$ 0.04 & 2.08 & 0.14 & 3.76 & 3\\
	29 & 130.580 & 1.654 & 23.97 $\pm$ 0.20 & 24.08 $\pm$ 0.03 & 0.51 & 0.59 & 0.49 & 4\\
	31 & 130.572 & 1.654 & 23.11 $\pm$ 0.12 & 23.46 $\pm$ 0.02 & 1.24 & 0.78 & 1.26 & 3\\
    \hline
  \end{tabular}
\end{table*}


\section{Results}
\label{sec:res}

\subsection{Cluster Membership}
\label{sec:clustmem}
\begin{figure*}
	\includegraphics[width=\textwidth]{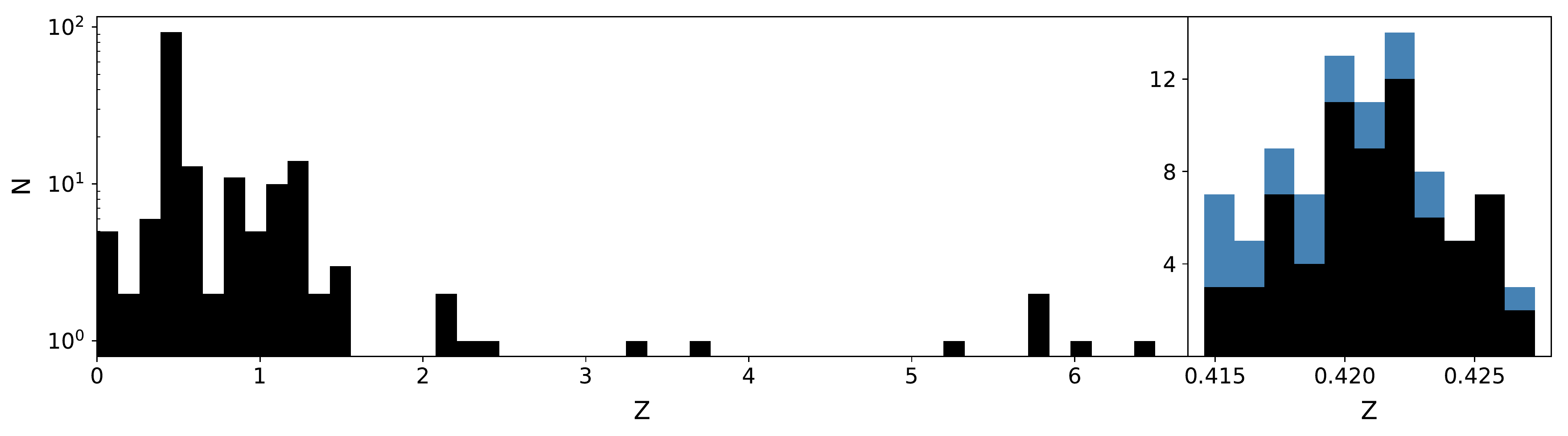}
    \caption{Left: Shows the distribution of spectroscopic redshifts within the observed MUSE field. Right: Shows only the objects within the redshifts of $0.41 \leq z \leq 0.43$ in which constraints have been placed on cluster membership. On the right, black represents absorption line galaxies and in blue are objects with emission line spectra.}
    \label{fig:groupz}
\end{figure*}

\begin{figure*}
	\includegraphics[width=\textwidth]{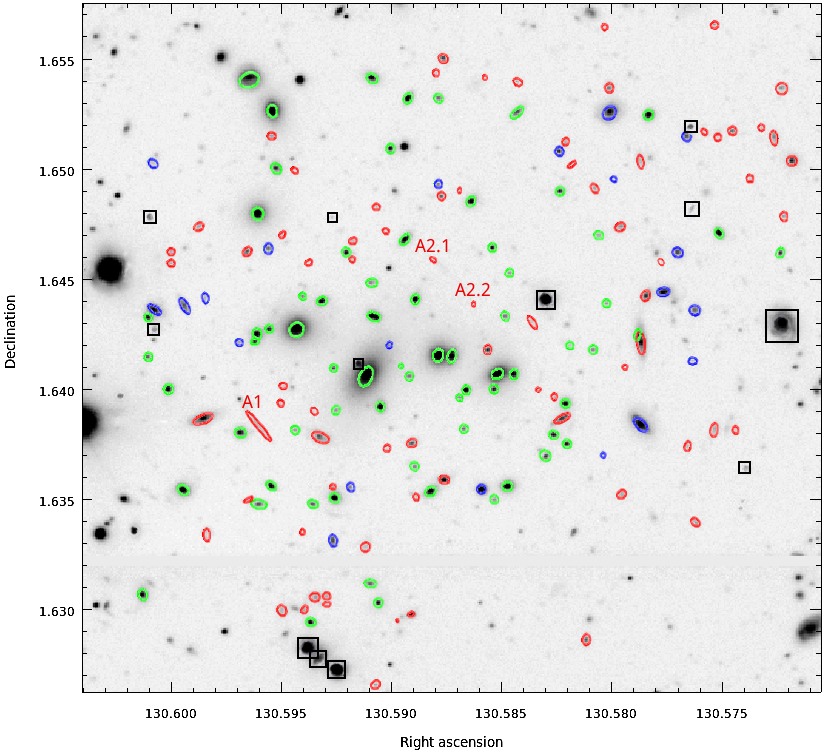}
    \caption{Spatial distribution of spectroscopically identified galaxies overlaid on the FORS2 r-band image. Black squares indicate foreground galaxies at z < 0.41, green and blue ellipses are absorption and emission line detected cluster members respectively, with redshifts $0.41 \leq z \leq 0.43$. Red ellipses show background galaxies with z > 0.43. All ellipses are scaled and positioned by SE\textsc{xtractor} parameters. A1 and A2 show the locations of the lensing arcs as discussed in Section~\ref{sec:lensmass}}
    \label{fig:regions}
\end{figure*}

\begin{figure}
	\includegraphics[width=\columnwidth]{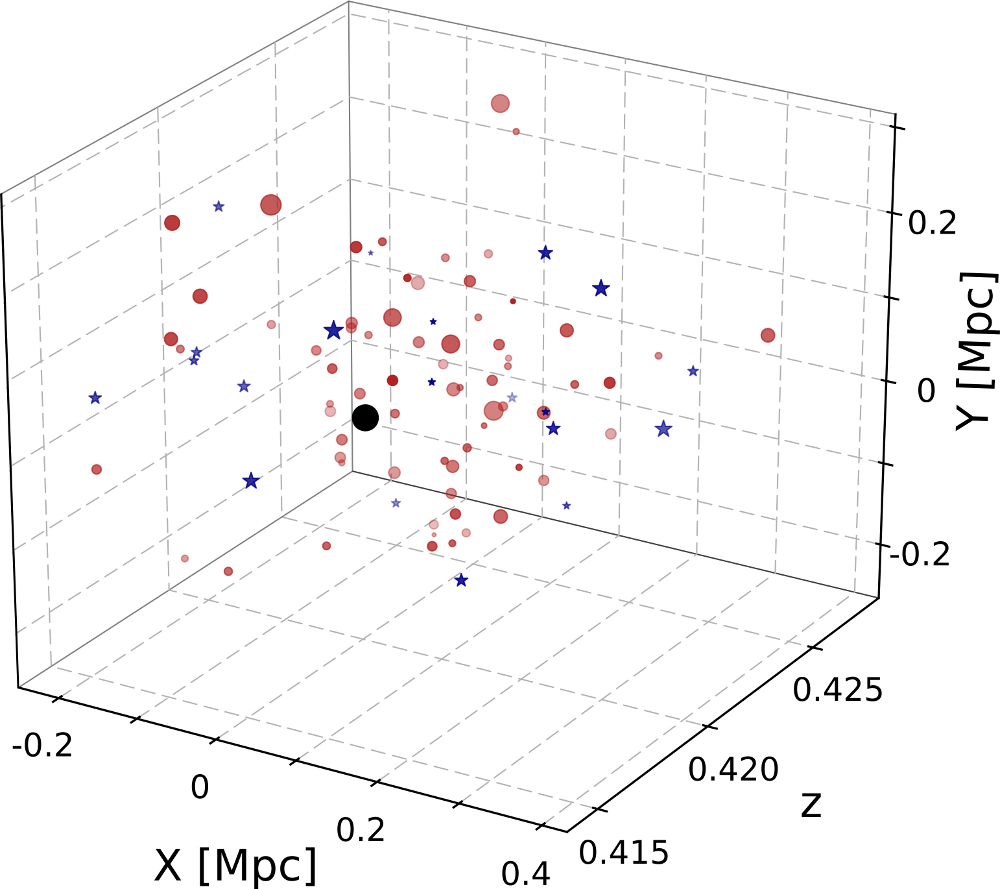}
    \caption{3D distribution of cluster members. The black circle represents the brightest cluster galaxy (BCG), red circles are galaxies fit with absorption dominated spectral templates, while blue stars are cluster members fit with emission line templates. Point sizes are scaled by object area using the SE\textsc{xtractor} parameter ISOAREA\_IMAGE.}
    \label{fig:3Ddist}
\end{figure}

Cluster members situated within the MUSE field were identified through the investigation of their spectroscopic redshift distribution, which can be seen in Figure~\ref{fig:groupz}. The tight clustering found around the previously quoted cluster redshift allowed for constraints on membership to be placed within 0.01 of the mean redshift of $z = 0.42$, in which a total of 89 galaxies are situated. The cluster sample predominantly consists of early type absorption galaxies with easily identifiable H and K spectral features, while 20 members show dominant [\ion{O}{ii}] and [\ion{O}{iii}] emission lines. The spatial distribution of spectroscopically identified cluster members can be seen in Figures~\ref{fig:regions} \&~\ref{fig:3Ddist}.

Further efforts were made in order to identify additional cluster members outside of the MUSE FoV using photometry derived from the FORS2 and IRAC data. Initially, a colour cut was performed, disregarding cluster members with $(g-r) > 2$, a linear fit was then applied to the spectroscopically identified red sequence in both (g-r) and (3.6$\mu m$-4.5$\mu m$), down to magnitude completeness limits of 25.5 and 19.8 mag, for r and 3.6$\mu m$ respectively. Any SE\textsc{xtractor} detected objects with colours within 1$\sigma$ of the fits were assigned as candidates for cluster membership. 

We investigate the blue cloud in a similar manner via spectroscopically identified emission line galaxies located within the cluster. As can be seen from the spatial distribution of cluster members in Figures~\ref{fig:regions} \&~\ref{fig:3Ddist}, there is a strong correlation between the cluster environment with galaxy spectral type. Galaxies identified via emission lines are situated at a mean radial distance of $\mathrm{\bar{r}}$ = 0.23 Mpc from the cluster centre, whereas absorption type members are found closer to the cluster core at $\mathrm{\bar{r}}$ = 0.16 Mpc. This truncation suggests that a higher fraction of cluster members beyond the MUSE FoV, are likely to be blue emission line galaxies. The combined analysis of both red and blue galaxies provides an additional 251 candidate cluster members.

In an attempt to further refine the colour-colour selection of potential cluster members, we utilise photometric redshift data recently made available in the first HSC public data release \citep{Aihara2017a}. Photometric redshifts within the HSC public data release are computed using several independent codes. Each of the codes are trained and verified by performing cross-validation techniques on a test sample, constructed from various spectroscopic and photometric sources \citep{Tanaka2017}. 

\begin{figure}
	\includegraphics[width=\columnwidth]{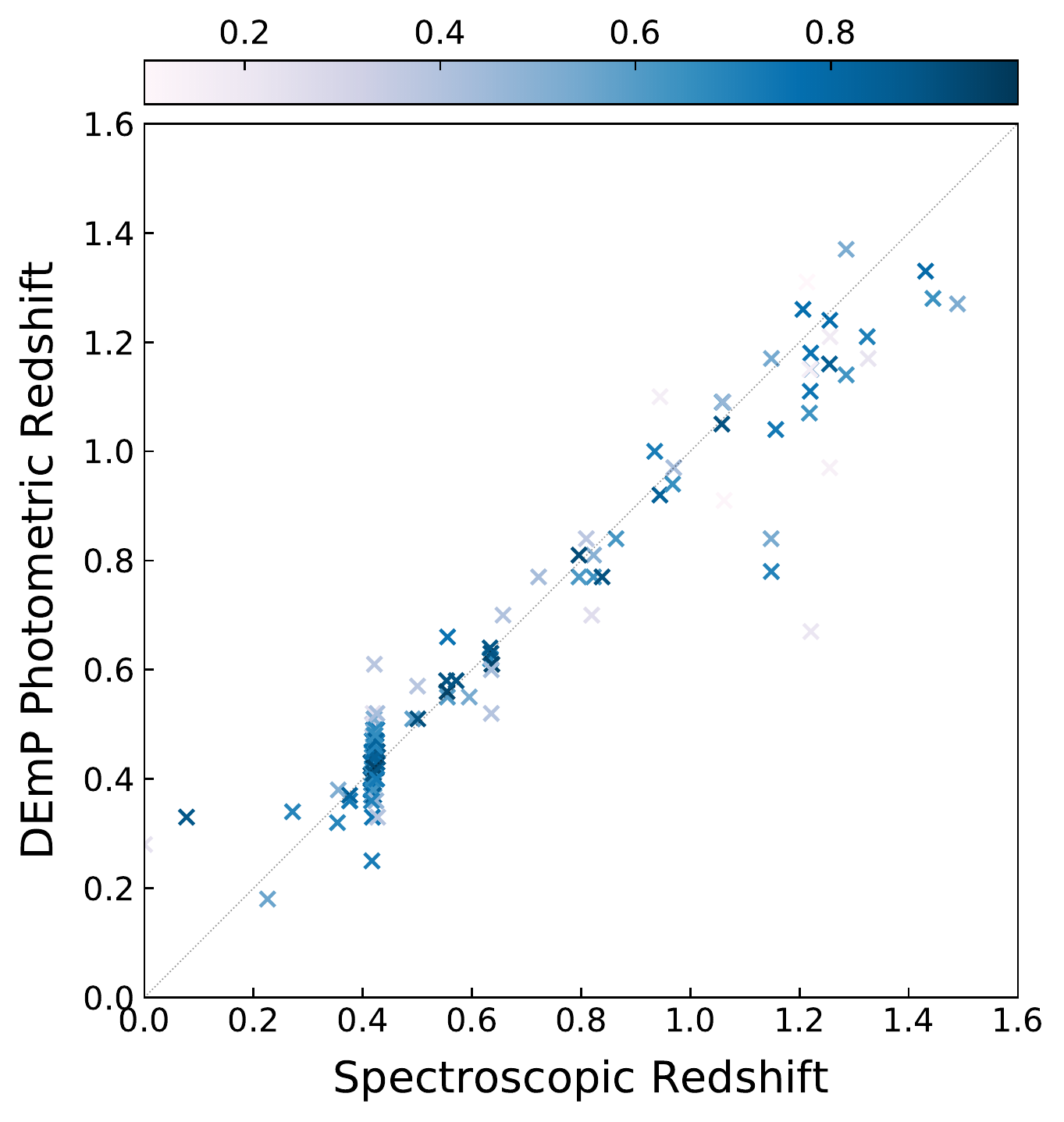}
    \caption{Comparison of spectroscopic redshifts found in this study to DEmP photometric redshifts from the first HSC public data release within the range of 0 < z < 1.6. Points are coloured by the photometric confidence value in which we require a value of > 0.5 for use in our analysis.}
    \label{fig:zvz}
\end{figure}

To identify which of the seven HSC photometric redshifts is best fit to our data, we perform a direct comparison to our spectroscopically identified cluster members. In order to achieve the most accurate results, we only include objects with a confidence level of > 0.5 in our analysis. We define outliers by the conventional definition of $|~\Delta z~| > 0.1$ and consider both the bias and the scatter of the residual $|~\Delta z~|~/~(1+z)$. Through this analysis, we find that the Direct Empirical Photometric code \citep[DEmP;][]{Hsieh2014} provides the best statistical fit to our spectroscopic redshifts, in which a comparison can be seen in Figure~\ref{fig:zvz}. We find an outlier rate of < 2\%, a bias of 0.015, scatter of 0.019. Over 80\% of the sample have a confidence value above the cut. 

We find similar results with the Ephor\footnote{Extended Photometric redshift (EPHOR) is a neural network photometric redshift code, see \citet{Tanaka2017} for details.}, and MLZ\footnote{Self-Organizing Map from the machine-learning photometric redshift package by \citet{Carrasco2014}.} photometric redshifts but with much lower sample numbers after the confidence cut ($\sim$70\% and $\sim$30\% respectively). We select all colour-colour identified galaxies with a DEmP photometric redshift of 0.42 $\pm$ 0.1 for a refined sample of 198 candidate cluster members.

Throughout this study, we use colour-colour selected members in our analysis only when detailed velocity information is not required. This includes projected cluster radii estimations, cluster centre and concentration analysis, as well as deriving the total stellar mass and luminosity function of the cluster.  

\subsection{Galaxy Populations}
\label{sec:gpop}

For all spectroscopically identified cluster members we measure (g-r) rest frame colours in which K-corrections were performed following methods outlined by \citet{Chilingarian2010}.

\begin{figure*}
	\includegraphics[width=\textwidth]{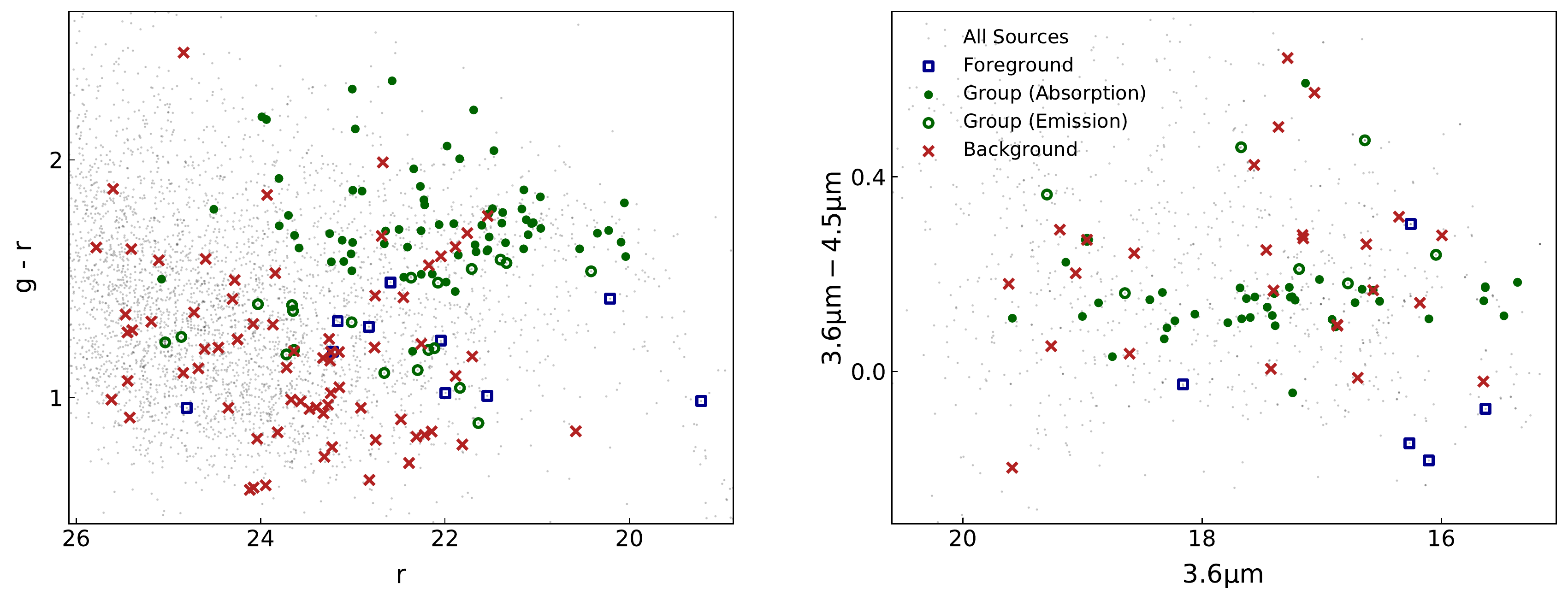}
    \caption{Colour-Magnitude relations of CLIO members, foreground and background galaxies. Left panel: Shows (g-r) colour vs r-band magnitude from FORS2 imaging, while right panel shows (3.6$\mu m$-4.5$\mu m$) vs 3.6$\mu m$ from IRAC. Green filled circles represent cluster galaxies which exhibit absorption spectra while green empty circles are emission line dominated cluster members. Blue plus symbols are foreground galaxies (z < 0.41). Light grey shows all, non-spectroscopically identified objects.}
    \label{fig:cmd}
\end{figure*}

A clear red sequence can be seen for the spectroscopically identified cluster members within both the FORS2 and IRAC data as shown in Figure~\ref{fig:cmd}. The cluster members have mean colours of (g-r) = 1.75 and (3.6$\mu m$-4.5$\mu m$) = 0.13 mag. The blue cloud, shown here through the distribution of emission galaxies, is observed in the FORS2 data to have a mean colour of (g-r) = 1.31 mag. The blue population is not prominent in the IRAC data due to the limited number of objects with corresponding (3.6$\mu m$-4.5$\mu m$) colours. 

It has been shown that there are tight correlations between optical and near-infrared galaxy colours with stellar mass-to-light ratios that can be modelled via a simple linear equation \citep{Sargent1974,Larson1978,Jablonka1992}. For a composite stellar population, the mass-to-light ratio depends on the initial mass function (IMF), the star formation history (SFH) and metallicity distribution. The work of \citet{Bell2001} show that through the use of stellar population synthesis (SPS) models, a reliable stellar mass can be estimated that is robust to metallicity and extinction effects. The most robust relations found are in the optical bands, where the age-metallicity degeneracy is beneficial, as it acts to tighten the relation. The work of \citet{Bell2001} also shows that dust extinction and reddening effects cancel each other out to first order approximation. In this study, we follow the methods outlined in \citet{Into2013} which takes into account the asymptotic giant branch phase in order to update colour-mass-to-light relations. We make use of the exponential models which more accurately mimics the photometric properties of the Hubble sequence and calculate stellar masses by:
\begin{equation}
	\rm{log_{10}}\left(\frac{M_*}{L_r}\right) = 1.373\left(g-r\right)-0.596.
\end{equation}    
The total stellar mass estimate of all combined cluster members is $\rm{log_{10}}\left(M_*/M_{\odot}\right) = (13.63 \pm 0.1~\rm{dex})$.

\subsection{Cluster Properties}

\subsubsection{Velocity Dispersion}
\label{sec:vd}
We measure the cluster velocity dispersion $\sigma$, with the robust gapper estimator method \citep{Beers1990}. The method uses weighted gaps in velocity space and is calculated as follows; firstly, recession velocities of all cluster members are ordered, gaps and weights between neighbouring pairs are then calculated via $g_{i}=v_{i+1}-v_{i}$ and $w_{i}=i(N-i)$ respectively, where $i=1,2,...,N-1$. The velocity dispersion can then be estimated by:
\begin{equation}
	\sigma = \frac{\sqrt{\pi}}{N(N-1)}\sum_{i=1}^{N-1}w_{i} g_{i}.
\end{equation}
To account for uncertainties in recession velocities, measurement errors are removed in quadrature as prescribed in \citet{Robotham2011}. Using the 89 spectroscopically confirmed cluster members, this method yields a velocity dispersion for CLIO of (619 $\pm$ 11) km s$^{-1}$. The total error is calculated via bootstrapping techniques. This value is as expected for a compact, high mass galaxy cluster \citep{AAQ2000} and comparable to the previous estimate of 633 km s$^{-1}$ from the G$^3$C catalogue. 

\subsubsection{Cluster Centre}
The projected cluster centre is calculated through an iterative process. An initial centre of light is defined from the r-band luminosities of the spectroscopically identified cluster members. The most distant galaxy in projected space is rejected, and the new centre of light is calculated. This process is iterated until only two galaxies remain, the brighter of which is taken as the cluster centre. The projected cluster centre was found to be located at the BCG, in agreement with the G$^3$Cv9 catalogue value at $\alpha =$ 08:42:21.88, $\delta =$ +01:38:26.11 (Galaxy ID: 323174). This iterative process was repeated to include the colour-colour selected galaxies to investigate if any bias is imposed by the positioning of the MUSE FOV, however, the resulting centre remained unchanged.

\subsubsection{Cluster Concentration}
\label{sec:conc}
Initial estimates of the cluster concentration of c = 6.32 are based on fitting the enclosed cluster mass to a NFW profile. Due to the availability in the G$^3$C, this is performed for the 50$^{\rm{th}}$ and 68$^{\rm{th}}$ percentile radii. These radii are calculated using the default quantile definition. Cluster members are sorted by ascending radial values from the projected cluster centre and a linear interpolation is performed between the bounding percentiles \citep{Robotham2011}.

Due to the limited extent of our spectroscopic coverage with MUSE, we identify cluster members out to only 0.4 Mpc from the cluster centre. This induces a strong radial bias, providing unreliable radius estimates. Even when including our colour-colour selected sample we find that our calculated radii are biased towards a lower value of $r_{50}$ = 0.27 Mpc (in comparison to the the G$^3$C value of 0.66 Mpc).

Here, we instead measure the concentration parameter by fitting a NFW profile to the projected number density of cluster members. As a function of radius, a NFW density profile is given by:
\begin{equation}
	\rho(r) = \frac{\rho_s}{\frac{r}{r_s}\left(1-\frac{r}{rs}\right)^2},
    \label{eq:nfwdensity}
\end{equation}
where r$_s$ is the scale radius, and $\rho_s = \delta_{s}\rho_c$ is a characteristic density, governed by the dimensionless concentration parameter $c_{200}$. The characteristic contrast density, $\delta_{s}$, is given by: 
\begin{equation}
	\delta_{s} = \frac{200}{3}\frac{c_{200}^3}{\ln(1+c_{200})-(c_{200}/1+c_{200})},
\end{equation}
and $\rho_c$ is the critical density of the universe at the cluster redshift. The scale radius $r_s$ describes where the density profile transitions from $\rho \propto r^{-1}$ to $\rho \propto r^{-3}$ and is of the form:
\begin{equation}
	\label{eq:conc}
	r_s = \frac{r_{200}}{c_{200}},
\end{equation}
where $r_{200}$ is the radius at which the density of the cluster is equal to 200$\rho_c$. Due to the limited spectral coverage, we refer to the methods presented in \citet{Carlberg1996} to measure $r_{200}$ via the cluster velocity dispersion by:   
\begin{equation}
	r_{200} = \frac{\sqrt{3}\sigma}{10H(z)}. 
\end{equation}
For $\sigma$ = 619 $\pm$ 11 km s$^{-1}$ and cluster redshift of 0.42, we calculate $r_{200}$ = 1.68 $\pm$ 0.03 Mpc.

With the scale radius as a free parameter, a $\chi^2$ minimization is performed, fitting a NFW profile to the clusters projected number density. We include both spectroscopically and colour-colour selected members to find a scale radius of r$_s$ = 0.22 $\pm$ 0.01. Equation~\ref{eq:conc} then yields, c$_{200}$ = 7.61 $\pm$ 0.43, a higher value than the initial estimate of 6.32. As the fitting procedures used in this work are not exactly the same as those from the previous concentration estimate, some discrepancy is expected between results. We believe that with the larger dataset available in this work, the concentration value obtained here is significantly more robust than the previous estimate.

We compare our results with lensing clusters selected from the CLASH survey. \citet{Merten2015} employs comprehensive lensing analysis to derive a concentration-mass (c-M) relation of 19 X-ray selected CLASH clusters between redshifts 0.19 and 0.89. The work reconstructs surface mass density profiles in order to derive NFW parameters. \citet{Merten2015} finds a concentration distributed around c$_{200} \sim$ 3.7 and an upper value of 4.7 for the MACS J1423+24 cluster. \citet{Siegel2016} combine gravitational lensing with multiwavelength analysis on 6 CLASH clusters and find good agreement with that of \citet{Merten2015}.

The NFW prescription predicts that at a given redshift, halo concentration is expected to decrease systematically with increasing mass \citep{Navarro1996}. As CLIO was selected because of its unique combination of high mass and concentration, extrapolations of c-M relations are likely to provide an underestimate of the total cluster mass. Adopting the method presented in \citet{Merten2015}, we find M$_{200} \approx 1.2 \times 10^{14}$ M$_\odot$. This is a factor of 5 difference from the G$^3$C initial estimate of 6 $\times 10^{14}$ M$_\odot$ and thus, likely cannot be used as a reliable mass estimate for this cluster.

Our measure concentration value of c$_{200}$ = 7.61 $\pm$ 0.43 confirms our initial assumptions of the high concentration of the CLIO cluster. Such a high concentration is also useful for producing strong lensing features. This can explain the presence of the giant arcs in this cluster, and may be indicative of a triaxial halo with its main axis along the line of sight.

\subsection{Luminosity Function}
\label{sec:lf}

Through the analysis of the luminosity function, we can obtain insight into the galaxy population of the cluster, unbiased by selection effects of limited spectroscopic coverage. For a previously unstudied cluster such as CLIO, the luminosity function can also provide information on the dynamical, and evolutionary state of the cluster, as well as serve as a comparison to other similar clusters.

The cluster luminosity function (LF) is defined as the statistical excess of galaxies along the line of sight of the cluster, with respect to a field sample. This implies that the background contribution along the cluster line of sight is equal to that of field samples. However, it is often the case that galaxy contaminants within the cluster or control field counts adversely affect the determination of the cluster LF.

We determine galaxy counts for both the cluster red sequence and blue cloud using our calibrated FORS2 g and r-band magnitudes. We select galaxies within a 3$\sigma$ fit of the spectroscopically identified red sequence (g-r) colour, down to the r-band photometric completeness limit of -16.5 Mag. We further select all galaxies with colour values bellow the fit as blue cloud members. Initially, we apply corrections to both data sets based on our MUSE spectroscopy in order to remove foreground source contamination in the cluster line of sight. The data are then binned in steps of 0.5 mag and a total cluster value is calculated through the addition of number counts within respective bins. Finally, we normalise number counts to units of Mpc$^{-2}$ mag$^{-1}$.

Without averaging over many clusters, statistical background subtraction can lead to large errors in the LF due to cosmic variance \citep[e.g.,][]{DePropris1999}. An alternative approach is to use spectroscopic data to efficiently estimate background counts without the need for statistical methods \citep{Muzzin2007}. However, due to limited spectral coverage we are unable to obtain robust background estimates in this study. Instead, we make use of HSC data to select a field sample from regions in close proximity to the cluster. This method has been shown to be comparable to that of the statistical methods \cite[see][]{Driver1998}. In order to avoid contamination from cluster member galaxies, we select by eye, 4 arbitrary field regions well outside the clusters virial radius. We define a minimum radial distance for field sample selection as twice that of r$_{200}$ from the cluster centre, corresponding to $\sim$ 3 Mpc. We follow the methods outlined above to derive a total, normalised galaxy count in each of the 4 fields. To further negate any statistical fluctuations within the individual fields, we compute a final, composite field sample by taking the average galaxy counts in each bin from the 4 normalised fields.

\begin{figure}
	\includegraphics[width=\columnwidth]{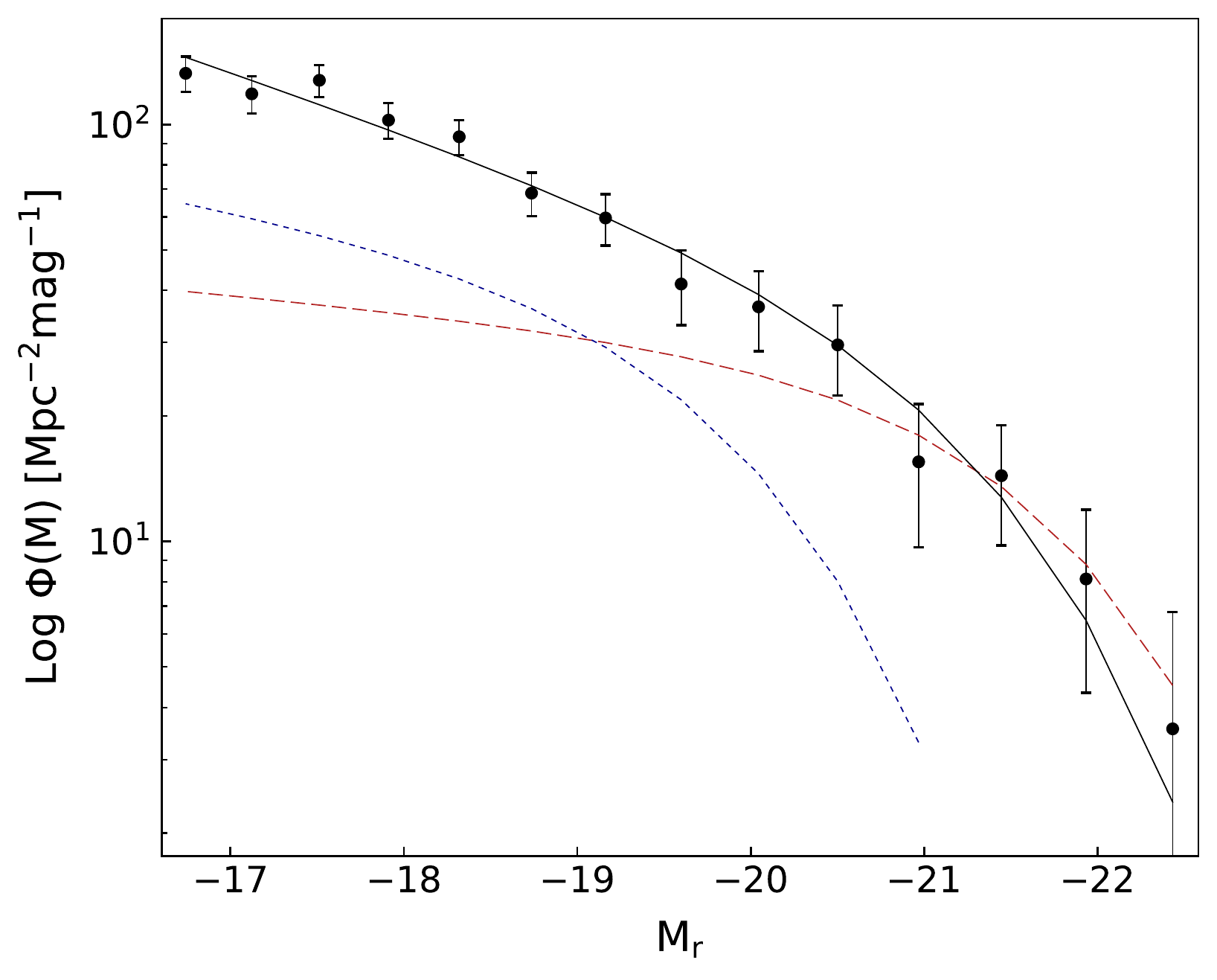}
    \caption{The total r-band, cluster galaxy count for CLIO, represented by black circles with uncertainties calculated considering both photometric and Poissonian errors. Data are split into 0.5 mag bins and are background corrected and normalised. Black solid line represents the best-fit Schechter function to the total cluster galaxy count. Red dashed, and blue dotted lines represent the best-fit Schechter function to the renormalised red sequence and blue cloud respectively. As the red and blue curves in this plot have been renormalised to match the total, the scales are not comparable and are shown for reference only.}
    \label{fig:lumfunc}
\end{figure}

We correct for background contamination by subtracting the control field number counts from the total cluster count in corresponding bins. The resulting LF is well described by a Schechter function with a slope of $\alpha = -1.35 \pm 0.05$ and a characteristic magnitude of $M^*_r = -21.54 \pm 0.04$ Mag. We calculate confidence intervals on the Schechter parameters via 1000 bootstrapping iterations in which we randomly re-sample 5\% of the data. The best fit Schechter function to the total LF is represented by the solid black line in Figure~\ref{fig:lumfunc}, along with renormalised fits to the red a blue galaxy distributions. 

When considering variations in magnitude range and cluster redshift, we compare $M^*_r$ and $\alpha$ values with composite cluster luminosity functions found in studies such as \cite{dePropris2003} and \cite{Christlein2003} ($M^*_r = −20.07,21.14$ and $\alpha = −1.28,-1.21$ respectively). These works (amongst others), find little variation in cluster luminosity functions with various cluster properties. Thus, it is not unexpected that we find consistent results here. At the faint end, we find the number density of blue galaxies increases beyond that of the red population. This can be expected for a cluster at this redshift due to the Butcher-Oemler Effect \citep{Butcher1984,Andreon2004}. Further, this supports our initial assumptions that our spectroscopic blue galaxy population is significantly truncated by the limited FoV of our MUSE observations.

\subsection{Intracluster light}
\label{sec:icl}

The ICL is the luminous stellar component, gravitationally bound to the cluster potential, but not to a particular galaxy. Thought to be formed via the tidal stripping of stars during hierarchical accretion histories of galaxy clusters and groups \citep{YenTing2004,Contini2014}. Thus, investigations of the ICL have provided detailed insight into the formation history of galaxy clusters.

\begin{figure*}
	\includegraphics[width=\textwidth]{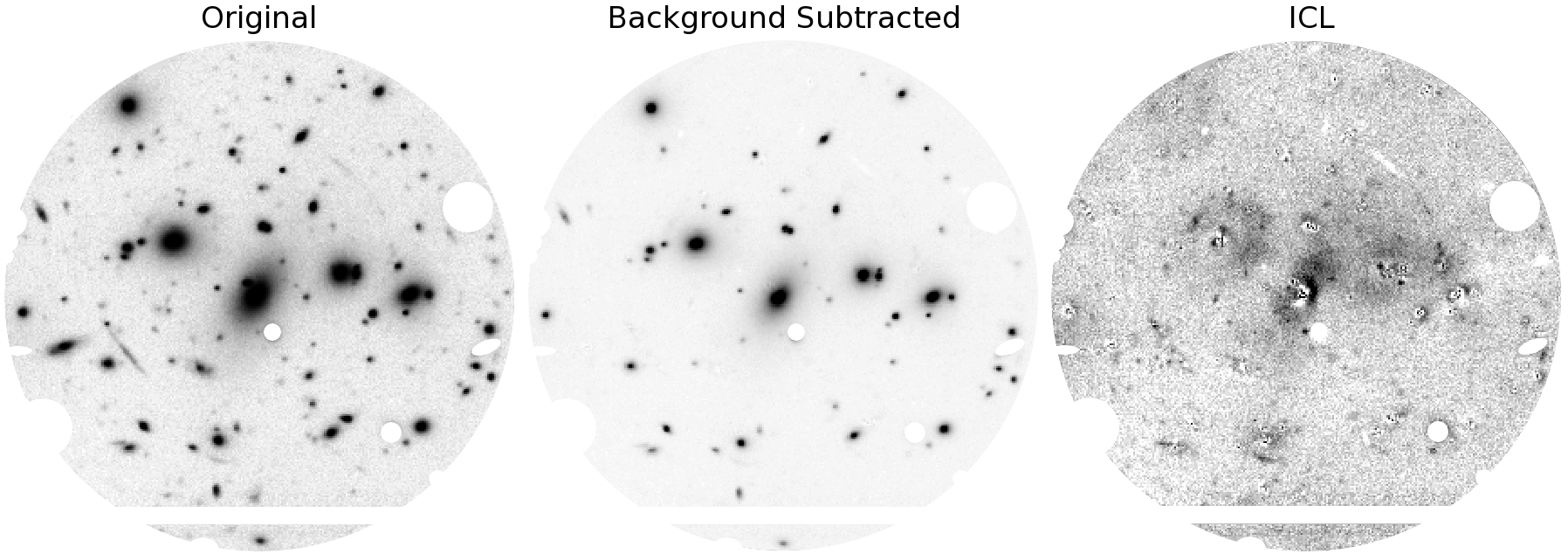}
    \caption{Visual representation of the full \textsc{Galfit} modelling procedure. Left: Shows the original FORS2 r-band image with initial masking applied. Centre: Shows a cluster only residual after subtraction of the composite foreground and background galaxy \textsc{Galfit} model. Right: Shows the residual ICL model after subtraction of the composite cluster model but before masking of the cluster member galaxy central regions.}
    \label{fig:fullmodels}
\end{figure*}

Working with a background subtracted r-band image, we measure the ICL as a fraction of the total cluster luminosity within the cluster core. We first mask everything outside of a 200 kpc radius from the BCG. We further mask stars and the intervening area between CCD's due to increased noise levels. This can be seen as the left image in Figure~\ref{fig:fullmodels}. Object detection is then performed by running SE\textsc{xtractor}, where we obtain x and y positions of objects as well as initial estimates of the effective radii R$_e$, position angles and axis ratios. Cluster members are then isolated from background and foreground galaxies using our detailed MUSE spectroscopy. Values obtained via SE\textsc{xtractor} are then used as initial parameters to model all non-cluster objects using \textsc{Galfit} \citep{Peng2010}. 

\begin{figure}
	\includegraphics[width=\columnwidth]{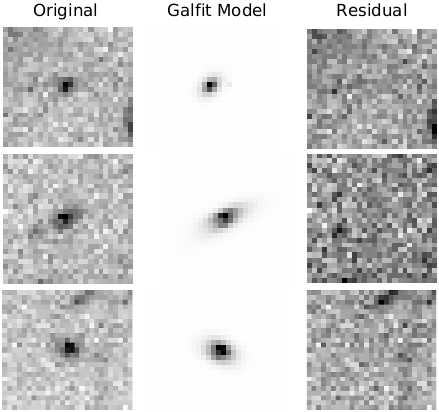}
    \caption{Illustration of \textsc{Galfit} modelling procedure. Left column shows postage stamp taken from FORS2 r-band image. Central column shows the \textsc{Galfit} S\'{e}rsic fit while the right column shows the residual after model subtraction.}
    \label{fig:models}
\end{figure}

To best estimate the local sky background, we create postage stamps centred on each object. When the galaxy can not be isolated successfully we fit multiple objects simultaneously. A single S\'{e}rsic profile is fit to each galaxy and the background is simultaneously modelled as a constant. A PSF estimated from a number of bright, non-saturated stars found within the cluster core is convolved with the profiles within \textsc{Galfit}. Each fit is manually checked through inspection of both the model and residual image, adjustments are made where necessary and in a number of cases an additional S\'{e}rsic profile was included. An illustration of the model fitting process is shown in Figure~\ref{fig:models}. Each model was then extracted in order to form a composite model of all background and foreground objects. This composite model is then applied to the masked r-band image resulting in a cluster member-only residual which can be seen in Figure~\ref{fig:fullmodels}, middle. 

We repeat this process, this time modelling the cluster members, a number of the larger galaxies are modelled with multiple S\'{e}rsic profiles in order to achieve the best fits. We further remove cluster members from the full image and thus leave only the residual ICL component as seen on the right of Figure~\ref{fig:fullmodels}. Finally, we apply additional masking to the central regions of some of the cluster members in order to avoid any residual contamination from \textsc{Galfit} modelling.

\begin{figure}
	\includegraphics[width=\columnwidth]{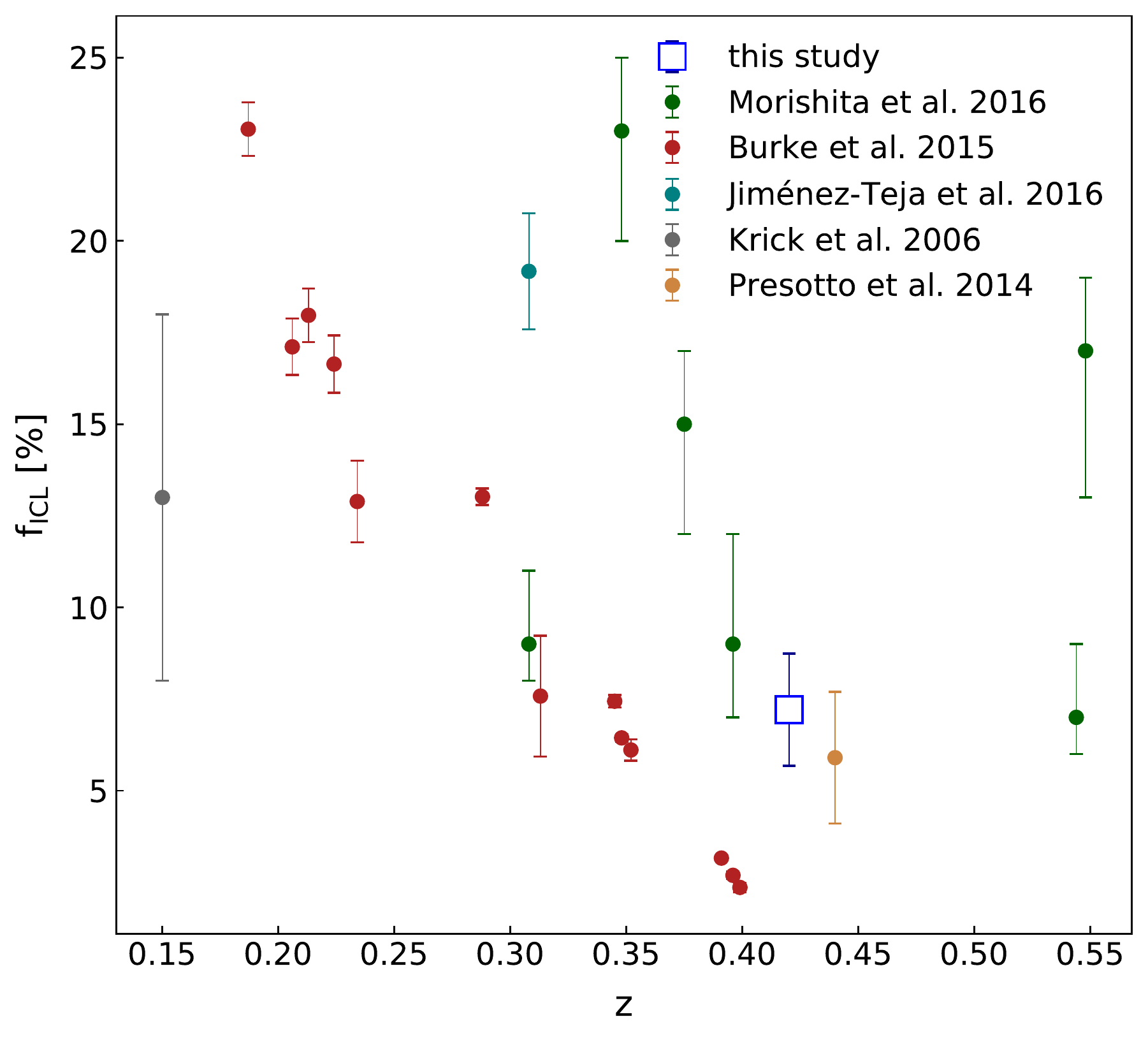}
    \caption{Comparison of galaxy cluster ICL fractions with cluster redshift. The ICL fraction obtained in this study is marked by an open blue square while other studies are marked with coloured points. As the works of \citet{Presotto2014} (yellow) and \citet{Morishita2016} (green) both use \textsc{Galfit} residual modelling, their results are likely the most consistent with those found within this study.}
    \label{fig:iclfraction}
\end{figure}

By comparing the total flux in the cluster member image to that of the ICL image, we can get a rough estimate of the ICL fraction of CLIO. This method provides an ICL stellar mass fraction of 7.21 $\pm$ 1.53\%, where error is estimated from the variance on the ICL flux ($f_{\textsc{icl}}$)
\begin{equation}
	\delta F_{\textsc{icl}} = \sqrt{\left(\frac{\sigma_{\textsc{icl}}}{f_{total}}\right)^2+\left(\frac{f_{\textsc{icl}}\sigma_{total}}{f_{total}^2}\right)^2},
\end{equation}
where $\sigma$ is the standard deviation of the flux. In Figure~\ref{fig:iclfraction}, we compare the ICL fraction of CLIO with the cluster ICL studies of \citet{Krick2006}, \citet{Presotto2014}, \citet{Burke2015}, \citet{Morishita2016} and \citet{Jimenez2016}. It should be noted that there is currently no standard method for the determination of ICL fractions, as such each study mentioned here employs various limiting radii and masking processes. The methods presented within this paper are motivated by the works of \citet{Presotto2014} and \citet{Morishita2016}. These studies use \textsc{Galfit} based residual modelling to eliminate contaminating galaxy light in order to determine ICL fractions. Whereas, \citet{Krick2006} adopts various modelling procedures to remove galactic light via masking. \citet{Jimenez2016} use methods based on Chebyshev Rational Functions \citep[CHEF;][]{Jimenez2012} while \citet{Burke2015} applies a surface brightness threshold. We primarily compare our results with those of \citet{Morishita2016}, whose study of six clusters between 0.30 < z < 0.55 find ICL fractions between 7 and 23\%. Our results here show that CLIO is consistent with the lower end of this distribution, matched only by the MACS1149 cluster at a higher redshift of z = 0.54. \citet{Presotto2014} finds the MACSJ1206.2-0847 cluster has a low ICL fraction of 5.9\%, comparable with that of CLIO. \citet{Burke2015} find values in the range of 2 - 3\% at a redshifts of z $\sim$ 0.4, however due to the variation in methods, a direct comparison with those found in this study is tenuous. We show here, that the low ICL fraction for the CLIO cluster is comparable with various clusters situated at similar redshifts. Thus, when considering the CLIO cluster for JWST lensing studies, this low ICL fraction is ideal in order to maximise the detection of faint lensed sources.

\begin{figure}
	\includegraphics[width=\columnwidth]{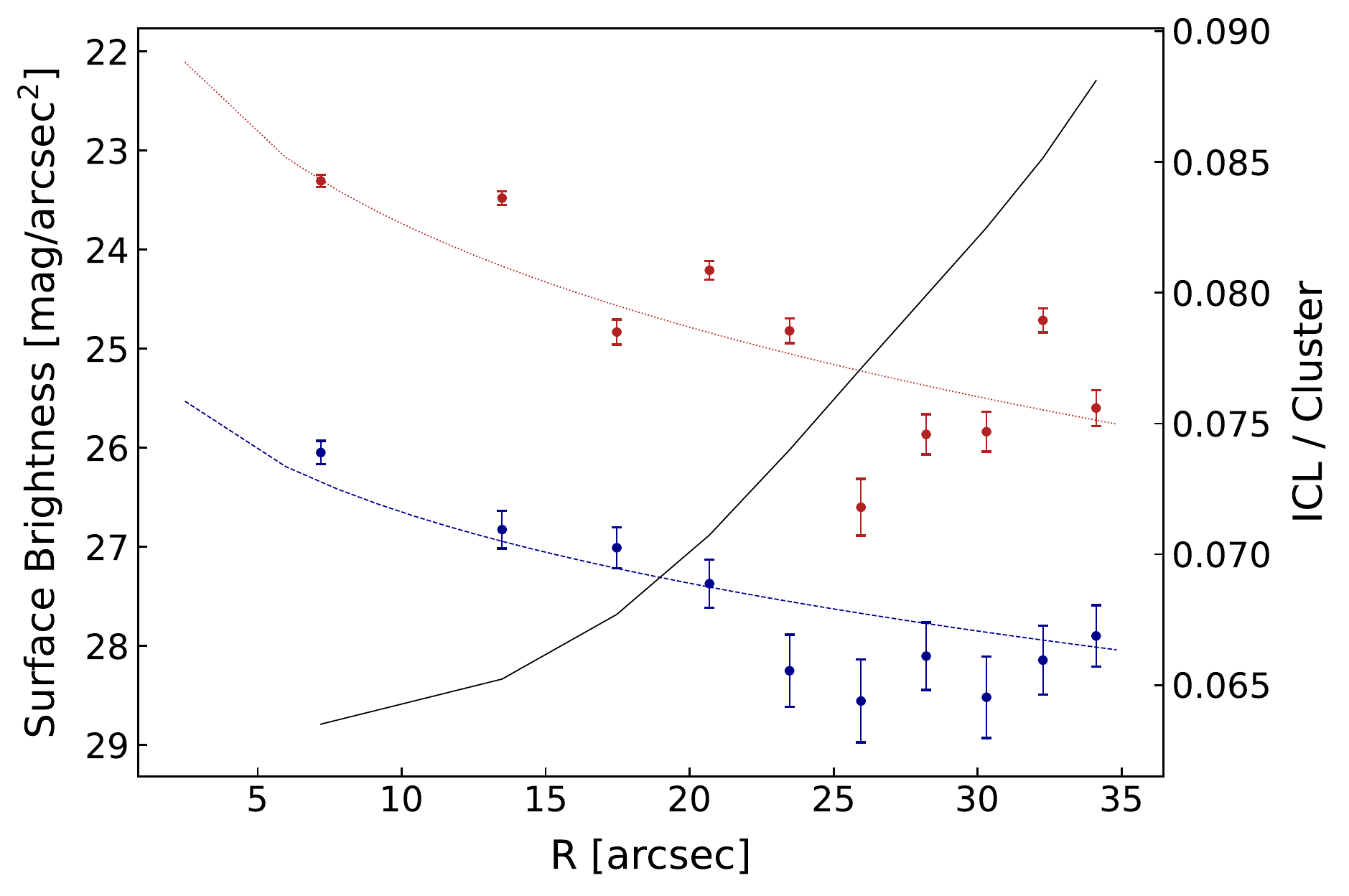}
    \caption{Radial surface brightness profile of the ICL (blue) and total cluster light (red), de Vaucouleurs profile fit to the ICL and cluster data are shown by the blue dashed and red dotted lines respectively. The dip seen at the outer edge of both the ICL and cluster surface brightness profiles is likely due to the initial masking process of stars and the CCD gap. The black solid line shows the ratio of the ICL to cluster light.}
    \label{fig:iclsbp}
\end{figure}

We measure the surface brightness profile of the ICL and total cluster light by splitting the image into radial annuli. We construct 25 radial annuli such that they contain equal areas of $\sim$ 150 arcsec$^2$. The recovered profiles can be seen in Figure~\ref{fig:iclsbp}. We find that there is a high degree of variation in the cluster radial profile due to radial positions of cluster members. In the outer regions, this is also enhanced by initial masking of stars and the CCD gap. Similarly, noise is induced in the ICL profile by initial masking effects, and also residuals from the galaxy modelling procedure.

To account for these fluctuations, we bin the data and fit a de Vaucouleurs profile to both the ICL, and cluster surface brightness profiles. The resulting ICL fraction ranges from as low as $\sim$ 6\% at the cluster centre where light is dominated by the BCG, up to $\sim$ 9\% towards the 200 kpc limits of this model. This trend is consistent with previous findings which have shown that the ICL fraction increases by around 2\% in the inner regions of clusters ($\lesssim$ 200 - 300 kpc) \citep{Rudick2011,Morishita2017}. Interestingly, the surface brightness of the ICL drops sharply at the position of the giant arcs ($\sim$ 20\arcsec). With a surface mass brightness of 26.63 and 27.71 mag/arcsec$^2$, internal and external of the radial arcs respectively, we see a drop of $\sim$ 1 mag. Minimal straylight contamination is crucial for the identification of faint lensed sources in the cluster centre. Thus, low ICL fractions such as those found for here are beneficial when considering target clusters for JWST observations.

\subsection{Mass Estimation}

\subsubsection{Dynamical Mass}
\label{sec:dynmass}
For a virialized system, the dynamical mass is expected to scale as $M \propto \sigma^2 r_{50}$, where $\sigma$ is calculated by considering only spectroscopically identified cluster members as detailed in Section~\ref{sec:vd}, and we take the G$^3$C value of $r_{50} =$ 0.66 Mpc. The dynamical mass can then be estimated by:
\begin{equation}
	\frac{M_{Dyn}}{M_{\odot}} = \frac{A}{G/M_{\odot}^{-1}km^2s^{-2}Mpc}\left(\frac{\sigma}{km s^{-1}}\right)^2\frac{r_{50}}{Mpc},
\end{equation}
where G = 4.301 $\times$ 10$^{-9}$ M$_{\odot}$ km$^2$ s$^{-2}$ Mpc is the gravitational constant and A is a scaling factor calculated to be 6.2 \citep[see][for more details]{Robotham2011}. The dynamical mass of CLIO, calculated via this method is M$_{\mathrm{dyn}} = (3.65 \pm 0.32) \times 10^{14}$ M$_\odot$. This value is lower than the catalogue value of $5.62 \times 10^{14}$ M$_\odot$ due to a combination of the refined velocity dispersion, and the significant increase in cluster multiplicity, which acts to decrease the scaling factor A from 9.1 to 6.2. This method is biased towards a lower mass for a number of reasons; the velocity dispersion is measured along the line of sight only, and, the intrinsic radius is likely to be larger than that of the projected radius due to a higher observed concentration of galaxies towards the cluster centre when viewed as a 2D distribution. Without a reliable $r_{50}$ value, we find that this method can not provide an accurate mass measurement. Instead, we employ various other techniques in order to obtain an estimate of the total cluster mass.

\subsubsection{NFW Mass}
\label{sec:nfwmass}
As we have spectroscopy of almost 90 cluster members, we make use of our robust velocity dispersion measurement (Section~\ref{sec:vd}) to estimate the total cluster mass. Assuming a spherical NFW mass distribution (as described in Section~\ref{sec:conc}) of the cluster, it is possible to obtain a total integrated mass. The mass enclosed within a given radius R is:
\begin{equation}
	M\left(<R\right) = \int_{0}^{R} 4 \pi r^2 \rho \left ( r \right ) dr,
\end{equation}
where $\rho(r)$ is given by Equation~\ref{eq:nfwdensity}. Taking the scale radius and characteristic density as described in Section~\ref{sec:conc}, we apply $r_{200}$ = 1.68 $\pm$ 0.03 Mpc as the limiting radius to obtain a total halo mass of: 
\begin{equation}
	M\left(<r_{200}\right) = 4\pi r_s^3\rho_s\left[\mathrm{ln}\left(1+c\right)-\frac{c}{1+c}\right] = \frac{800\pi}{3} r_{200}^3\rho_c.
\end{equation}
This is also know as the spherical overdensity mass. As our calculation for $r_{200}$ depends only on the projected velocity dispersion and cluster redshift, the radial biases previously encountered (Section~\ref{sec:conc}) can be avoided. This provides a mass estimate of M$_{200} = (4.49 \pm 0.25) \times 10^{14}$ M$_\odot$. This method, is however limited by the assumption that the cluster is well described by a spherical profile. This is unlikely to be accurate, even within the limited MUSE field it can be seen that there is a excess of large cluster member galaxies extending north-east out from the BCG. Despite this, this method provides a good initial estimate for the total cluster mass.

\subsubsection{Stellar Hydrodynamical Mass}
\begin{figure}
	\includegraphics[width=\columnwidth]{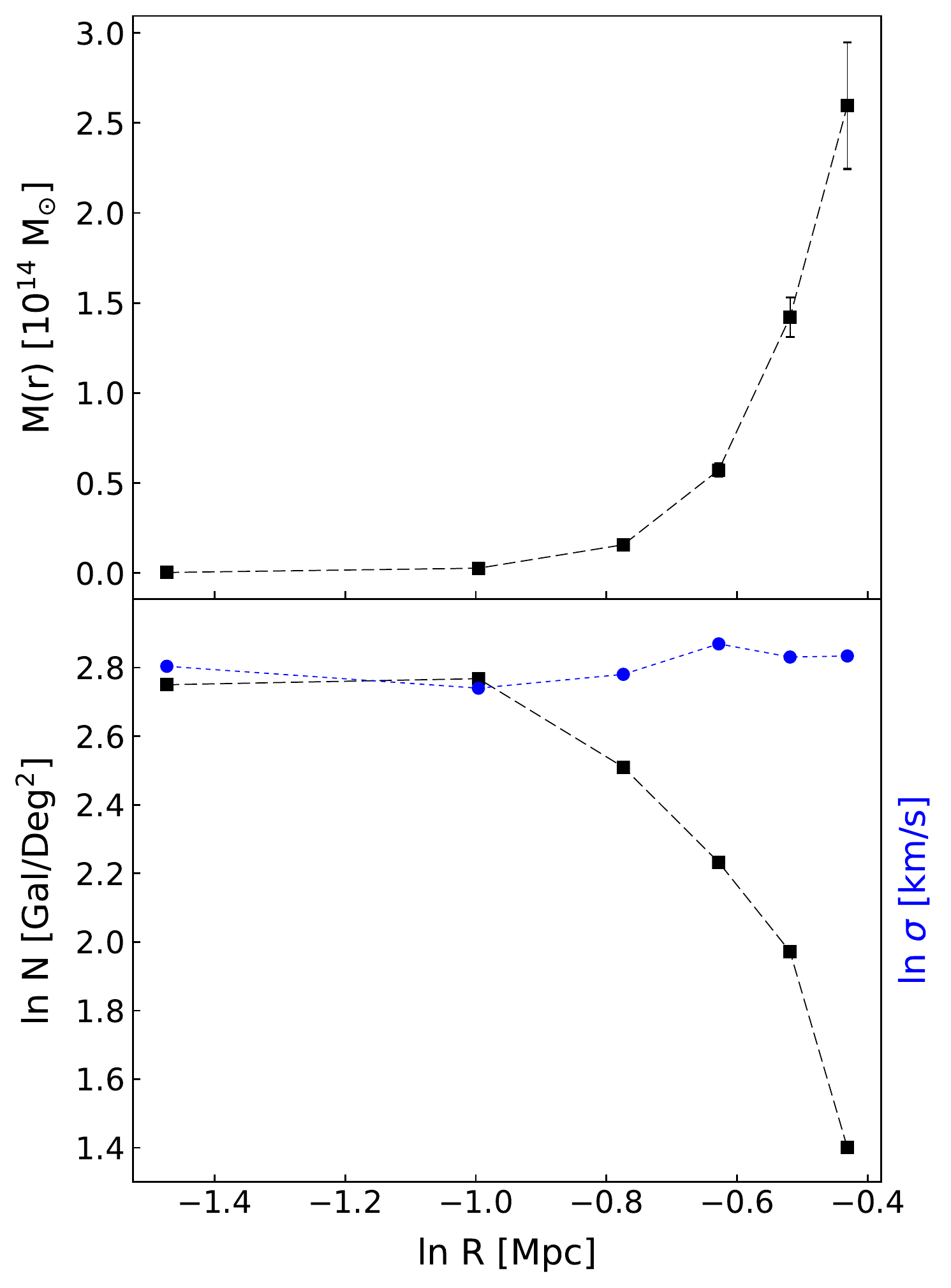}
    \caption{Top: Shows the radial mass profile of CLIO calculated via Equation~\ref{eq:rmp} with a total mass estimate of $(4.18 \pm 0.51) \times 10^{14}$ M$_\odot$. Bottom: Blue cicles represent the natural logarithm (ln) of the radial velocity dispersion ($\sigma_r^2$) in km s$^{-1}$, while black squares are the natural logarithm of the radial number density of cluster members ($v$) in units of galaxies per square degree.}
    \label{fig:massprofile}
\end{figure}

The cluster mass of CLIO can be further investigated through the use of the Jeans equation. This relates spatial and velocity distributions of particles moving in spherical orbits, to the systems mass profile \citep{Carlberg1997,GalacticDynamics}
\begin{equation}
	M\left(r\right) = -\frac{\sigma_r^2r}{G}\left[\frac{\mathrm{d~}ln(\sigma_r^2)}{\mathrm{d~}ln(r)}+\frac{\mathrm{d~}ln(v)}{\mathrm{d~}ln(r)}+\beta\right].
    \label{eq:rmp}
\end{equation}
We will refer to this as the stellar hydrodynamical mass, where $\sigma_r^2$ is the radial velocity dispersion, $v$ is the radial number density profile, and $\beta = 1 - v_\theta^2/v_r^2$ is the anisotropy parameter. It is important to bear in mind that it is impossible to know the true value of $\beta$ and the equation is not likely to provide physical results for highly radial orbits. From our investigations into the luminosity function (Section~\ref{sec:lf}) and ICL fraction (Section~\ref{sec:icl}) of the cluster, we find an excess of blue galaxies and low rates of ICL. Thus, we can assume that the cluster is still in a dynamical state. This means that a significant fraction of cluster members are likely to have more radial, rather than spherical orbits. However, if we make a simple assumption that the velocity ellipsoids follow spherical orbits, then $\beta = 0$, and a mass of $(4.18 \pm 0.51) \times 10^{14}$ M$_\odot$ is found. This full mass profile can be seen in Figure~\ref{fig:massprofile}. It is worth noting that this mass is only constrained within 0.4 Mpc ($\sim$ 0.25r$_{200}$). With such high concentrations, a high fraction of the cluster mass is likely to be contained within its core, this value may initially seem like an overestimate based on our previous findings of M$_{200} = (4.49 \pm 0.25) \times 10^{14}$ M$_\odot$. However, if we reconsider the assumption that $\beta = 0$, the enclosed mass decreases down to M$(<0.4 $Mpc$) = 3.51 \times 10^{14}$ M$_\odot$ for $\beta = 1$. In reality, the true value of $\beta$ is likely to fall somewhere between 0 and 1.

We estimate the dark matter halo mass fraction of the cluster by comparing the cluster mass, with that of the stellar masses calculated in Section~\ref{sec:gpop}. By adopting the stellar hydrodynamical mass calculated in this section we find a dark matter halo mass fraction of 91\%. Comparisons against the dynamical (\ref{sec:dynmass}) and NFW mass (\ref{sec:nfwmass}), provide estimates of 88\% and 90\% respectively.

\subsection{Preliminary Lensing Models}

We attempt to locate potential multiply imaged galaxies in order to build a robust strong-lensing model. Initially, we search for galaxies that are closely situated in colour-redshift space, where we find 14 potential multiple-image groups consisting of a total of 43 individual images. Further investigation is then performed by stacking the spectra of each galaxy in the group to identify shared spectral features. Although a number of these did show promise, upon further investigation into the spatial locations of the galaxies, we found that none of them were reliable enough to be used in the initial modelling. 

We identified multiple lensing arclets within the optical imaging, and made efforts to extract quality spectra from the MUSE datacube. We find two sets of arclets at roughly the same radial distance from the cluster centre (lower left and upper right of the cluster core as seen by points A1 and A2 respectively in Figure~\ref{fig:regions}) indicating the source galaxies are likely situated within a similar redshift range. At the lower left we find an extended arclet, the spectra presents no strong emission or absorption features, however we fit suspected \ion{C}{iii}, \ion{C}{iv} and \ion{Mg}{ii} features to find a preliminary redshift of $z = 2.37$. The upper right arc is less luminous and appears to have multiple components, indicating this is a fold arc with a single source, or individual images of two galaxies situated in close proximity in the source plane. Again, the spectra of which shows no strong features but we fit a preliminary redshift of $z = 2.18$ to both components. There is a small shift between fits of the two components, indicating that the images may be from separate, but closely situated source galaxies, however more accurate spectra is required to verify this. The redshifts of these arclets, although of lower quality (Q = 2) are included in our redshift catalogue, as they have been used to constrain our preliminary lensing models. 

\subsubsection{Light-Traces-Mass modelling}
\label{sec:lensmass}

Given that no obvious multiple-imaged galaxies were detected in the data, a first, rough strong-lensing strength estimate for the cluster was obtained with the automated Light-Traces-Mass code \citep[LTM;][]{Zitrin2012a}. This method relies on the simple assumption that light traces both the stellar and roughly the dark mass within the cluster. The LTM model is particularly applicable here as it can be self-calibrated on other lensing clusters with known images. This yields a good initial estimate for the mass distribution, even without multiple image inputs \citep{Zitrin2012a}. While the mass distribution and locations of critical curves have been found to be well constrained by the cluster light distribution, the mass profile remains uncertain. Without available multiple images, extrapolation of the mass profile out to M$_{200}$ is unreliable. Thus, the most secure method of mass estimation is to compare the Einstein radii to a distribution of clusters. Using a sample of carefully analysed clusters from the GMBCG catalogue \citep{Hao2010}, Zitrin et al. developed a LTM scaling relation that predicts the location of the critical curves from the luminosity distribution of cluster members, which in turn can be used to estimate M$_{200}$ and a corresponding uncertainty.

Due to the lack of multiple images, we initially use the self-calibrating LTM model (hereafter, LTMv1). For this, we use the FORS2 red sequence member galaxy list, which includes both spectroscopic and colour-colour identified member galaxies. This method yields a relatively small effective Einstein radius of $8.4\arcsec\pm20\%$, enclosing a mass of $2.02 \times 10^{13}$ M$_{\odot}$. The small Einstein radius is in agreement with the lack of multiply imaged galaxies identified. Comparing to Weak Lensing results form the CLASH survey \citep{Umetsu2016}, this Einstein radius typically corresponds to clusters with M$_{200}$ of about few $10^{14}$ M$_\odot$.
  
In addition, we construct a secondary model from the \citet{Zitrin2012a} automated method (we refer to this model as LTMv2). This time, the critical curves are scaled by hand to match the radius in which the arcs are seen, providing an effective upper limit on the Einstein radius and thus, cluster mass. This model yields an effective Einstein radius of $r_E \sim 15\arcsec$ enclosing a mass of $\sim 4.5 \times 10^{13}$ M$_{\odot}$. Following the Umetsu results mentioned above, this radius typically corresponds to a higher mass cluster than our first model estimate, yielding an upper limit on the total mass of M$_{200} \sim 10^{15}$ M$_{\odot}$.

\subsubsection{Weak \& Strong Lensing Analysis Package}

We perform complimentary lensing mass model reconstruction using the Weak \& Strong Lensing Analysis Package \citep[WSLAP+;][]{Diego2005,Diego2016}. WSLAP+ is a free-form method used to model gravitational lenses using a combination of week and strong lensing data. The mass in the lens plane is modelled as a combination of a diffuse, and compact component. The diffuse component is a superposition of Gaussian functions located at a distribution of regular or adaptive grid points. The compact mass accounts for the baryonic and dark components associated to the elliptical-type member galaxies selected from the cluster red sequence. The distribution of mass is assumed to trace the light of these compact galaxy components. A detailed description of the code and the different improvements implemented can be found in the papers \citet{Diego2005}, \citet{Diego2007}, \citet{Sendra2014} and \citet{Diego2016}. 

We employ the spatial distribution of cluster member galaxies in combination with arclet positions, their extent, and redshifts as constraints to derive a strong lensing mass model using the WSLAP+ package. The arclets are mapped in their entirety (not just their centroids) and included as constraints. Instead of assuming the centroid positions of the arclets (1 point for the arc in the lower-left side of the field of view at $z_1 = 2.37$, and 2 points for the two arclets in the upper-right side of the field of view at $z_2 = 2.18$) as it is usually done in lensing reconstruction methods, we consider 11, and 17 points for the arcs at $z_s = 2.37$ and $z_s = 2.18$ respectively. A table of central arc positions, along with an illustration of the points used, can be found in Appendix~\ref{apx:lensing}. These points are roughly equally spaced with a separation of $\approx 0.7''$ between them. The points do not correspond to particular features in the observed arcs but rather they trace the shape and extension of the arcs. The algorithm exploits the extension of the arclets by requiring that the lens model \textit{focuses} the extended arcs into small compact regions in the source plane. This approach is very useful when the number of constraints is scarce but there are giant arcs in the lens plane.
  
Also, incorporating the member galaxies in the lens model acts as an \textit{anchor}, constraining the range of possible solutions and reducing the risk of a bias due to the minimization being performed in the source plane. The algorithm determines the optimal distribution of mass by minimizing the lens equation under the additional constraint that the mass in each cell must be positive. For this particular cluster, we use a regular configuration with $8 \times 8$ grid points (or Gaussians) to fit the diffuse component. All member galaxies are assumed to have the same mass-to-light ratio which is determined as part of the optimization process.
 
\begin{figure}
	\includegraphics[width=\columnwidth]{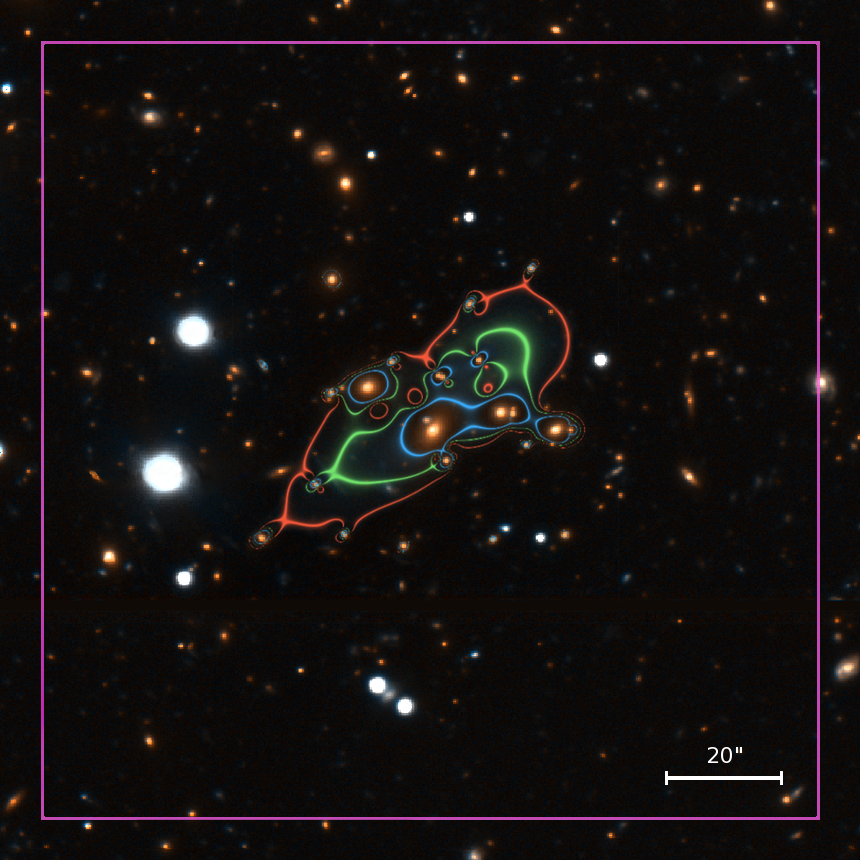}
    \caption{False colour image constructed with g and r-band FORS2 data showing the central region of the CLIO cluster. Critical curves for the preliminary lensing model are shown in blue, green and red. The blue line shows the tangential critical curve for a redshift of z = 1.2, where there is a peak in the redshift distribution. Green, represents the critical curve for the estimated arclet redshift of z = 2.2. Finally, the red line is the z = 6 critical curve. The JWST NIRCam Module A FoV is overlaid as the magenta box.}
    \label{fig:lensmodel}
\end{figure}

This preliminary model, which can be seen in Figure~\ref{fig:lensmodel}, provides no further constraints at this time as no counterimages are identified using our current data. Without weak lensing analysis, or additional strongly lensed systems, the WSLAP+ mass model reconstruction only constrains the central part of the galaxy cluster, out to a radius of $\sim$ 35 arcseconds. We extrapolate the integrated mass profile out to r$_{200}$ by fitting a NFW to estimate a total cluster mass. While fitting well to the density profile within the $\sim$ 35, even slight variations in the concentration of the NFW profile provide vastly different results when extrapolated out to r$_{200}$. To provide the most robust estimate, we use our previously calculated concentration (Section~\ref{sec:conc}) of 7.61 to obtain a total cluster mass of $\sim 6 \times 10^{14}$ M$_{\odot}$. It is worth noting that strong lensing only extrapolations tend to overestimate total cluster mass, since clusters are not axially symmetric and often present complex substructure.

\begin{figure}
	\includegraphics[width=\columnwidth]{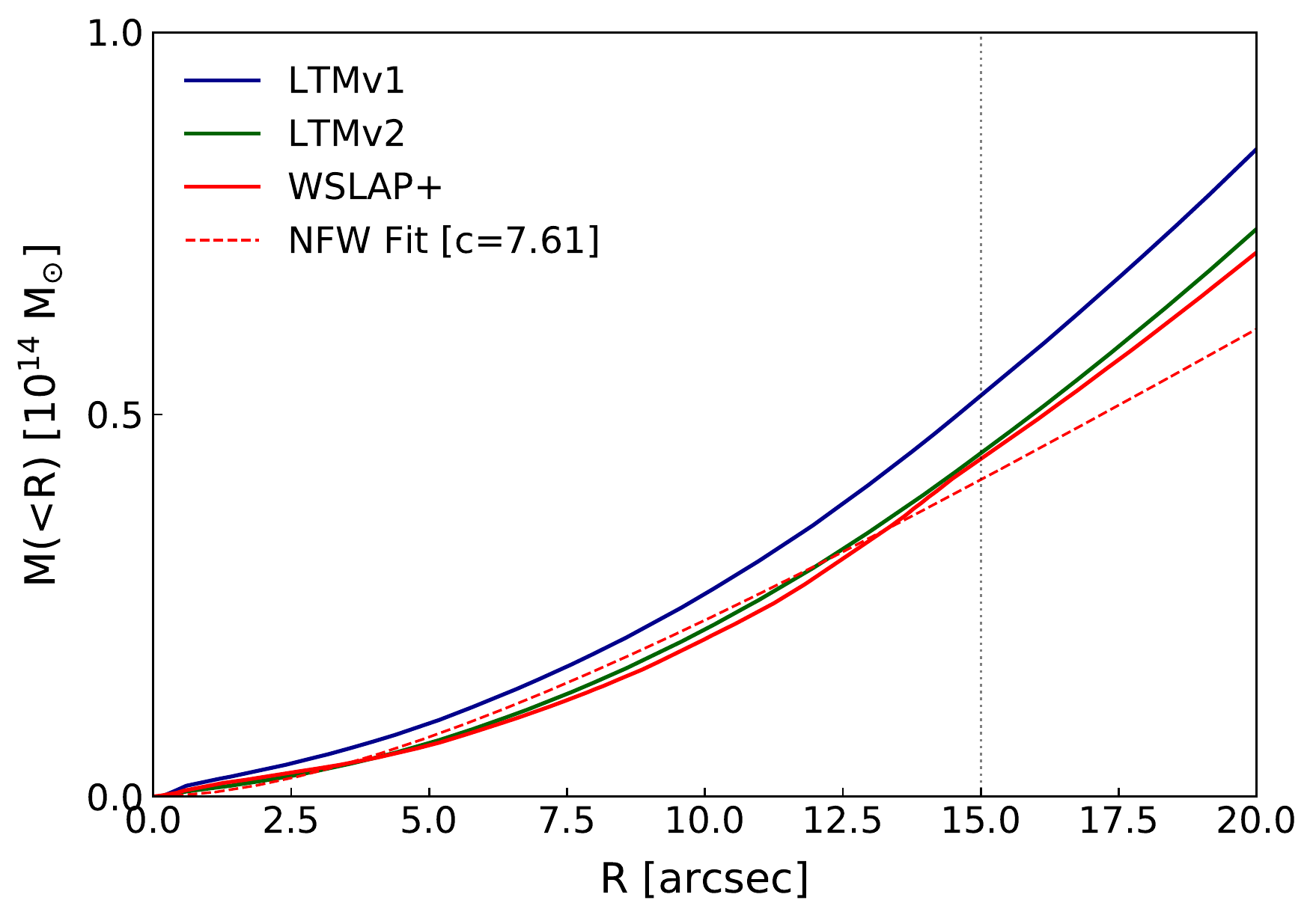}
    \caption{Enclosed mass profiles derived from both LTM, and WSLAP+ lensing models. The blue line shows the self-calibrated LTMv1 model which provides an Einstein radius of 8.4\arcsec, which encloses a mass of $1.29 \times 10^{13}$ M$_{\odot}$. The green line shows the LTMv2 model, calibrated to the observed lensing arcs, providing and Einstein Radius of 15\arcsec (grey dotted line), enclosing a mass of $4.33 \times 10^{13}$ M$_{\odot}$. In red we see that the WSLAP+ model, also calibrated to the observed arcs, provides a consistent mass profile within the Einstein radius, enclosing a mass of $4.42 \times 10^{13}$ M$_{\odot}$. The red dashed line shows the NFW profile used for the WSLAP+ mass estimate. It can be seen that while the models agree well within the Einstein radius, they deviate significantly when extended to larger radii.}
    \label{fig:lensprofiles}
\end{figure}

Through the use of observed lensing arcs, we are only able to constrain the mass enclosed within the Einstein radius of the lens, and not the full mass profile. While the LTM method constructs a projected mass distribution based on the observed light, WSLAP+ does not make the same assumption. Thus, the density profiles of the models are intrinsically different, even if the integrated mass within the Einstein radius may agree. In Figure~\ref{fig:lensprofiles}, we show that the mass profiles calibrated using arc positions (LTMv2 and WSLAP+) are consistent within the Einstein radius of the lens. However, it can be seen that these profiles differ significantly beyond 15\arcsec and hence, extrapolations out to larger radii can result in significant differences in calculated enclosed masses. The addition of new arcs at larger radial distances, or weak lensing data will allow further constraints on the density profiles to reduce uncertainty.

The extension of the critical curve predicted from our model (at z $\sim$ 3) is able to magnify an area smaller, but still comparable, to that of the powerful lenses. Since (to first order) the area above a given magnification in the source plane is proportional to the same area in the image plane divided by the magnification, CLIO is able to magnify (above a given magnification) about half the area when compared with the most powerful lenses. This is beneficial for JWST NIRCam observations due to the limited FoV pointing's. Given the fact that the main factor determining the number of observed high redshift galaxies around a lens is cosmic variance (i.e, the number of high redshift galaxies that fall in the footprint of the high magnification region in the source plane), it would not be surprising if high redshift galaxies are found behind CLIO when observed with powerful IR telescopes like JWST.


\section{Summary}
\label{sec:summary}

We have presented the first measurements and detailed study of the compact galaxy cluster CLIO. We perform standard data reduction procedures on dual band FORS2 optical imaging and multiple channel IRAC imaging. We perform object detection on the r-band image in order to create spatial profiles for extraction of weighted spectra from the MUSE IFU datacube. Redshift analysis is performed using a customised version of the cross correlation \textsc{marz} code, results are checked against findings from python based emission line detection packaged \textsc{muselet}. We identify 89 cluster members at redshifts $z = 0.42 \pm 0.01$ and a further 75 background objects out to z = 6.49.  

Photometry is measured for all detected objects in the FORS2 field and we find corresponding galaxies in the IRAC data. After applying k-corrections and considering extinction effects we calculate (g-r) and (3.6$\mu m$-4.5$\mu m$) colours. We then use a colour to mass-to-light relation to estimate stellar masses of all spectroscopically identified cluster members. Additionally, we use galaxy colours and HSC photometric redshifts to measure the cluster red sequence and search for additional cluster candidates outside of the spectroscopically covered field, yielding a total of 198 additional galaxies.

We investigate cluster properties to update initial G$^3$C estimates. Using the 89 spectroscopically identified cluster members, we calculate a velocity dispersion of $\sigma$ = (619 $\pm$ 11) km s$^{-1}$ using the robust gapper estimator. We use an iterative process to find the projected cluster centre located at the BCG and is not sensitive to interlopers. 

There is a radial bias induced by the limited MUSE coverage, so adopt a velocity dispersion based estimator for r$_{200}$. We calculate a concentration value of c = 7.61 $\pm$ 0.43 by fitting a NFW profile to the clusters projected number density. This high concentration confirms initial assumptions made during cluster selection. Additionally, a concentration value such as this is thought to explain the presence of the strong lensing features observed, and may be indicative of a triaxial halo.

Through the investigation of the luminosity function we find an excess of faint blue galaxies when compared to lower redshift samples. This indicates that the cluster is yet to be viralised. Further, we investigate the ICL fraction of CLIO through parametric profile fitting using the \textsc{Galfit} software package. Through the analysis of object subtracted residual images we find an ICL fraction of 7.21 $\pm$ 1.53\%, which is in the lower range of ICL measurements at similar redshifts. This low ICL fraction supports both the high concentration of this cluster, and the excess of the faint blue population. Further, it suggests that the ICL in CLIO is still in development, making it an interesting candidate for JWST lensing studies.

We find that due to the high concentration of the cluster, any mass estimates based on radial measurements prove unreliable. Instead we adopt two separate estimates based on velocity dispersion. We first assume the mass is distributed as a NFW profile and choose the enclosing radius as r$_{200}$ to achieve a purely velocity-dispersion based estimation of M$_{200} = (4.49 \pm 0.25) \times 10^{14}$ M$_{\odot}$. Secondly, we apply Jeans analysis to estimate the stellar hydrodynamical mass contained within 0.4 Mpc to be between M$(<0.4 $Mpc$) = (3.51 \pm 0.39) \times 10^{14}$ M$_{\odot}$ and $(4.18 \pm 0.51) \times 10^{14}$ M$_{\odot}$, for $\beta$ = 1 and 0 respectively.

Finally, analysis of galaxy spectra initially provides a number of potential multiply imaged galaxy candidates, however, we cannot confirm these with our existing data. Despite this, we use the automated Light-Traces-Mass and WSLAP+ methods to provide a rough strong-lensing estimate. With both models calibrated to the positions of the lensing arcs, we constrain the Einstein radius to be at r$_E \sim$ 15\arcsec. LTM and WSLAP+ models calibrated in this way provide consistent mass profiles within r$_E$. They are however unreliable when extrapolated to larger radial distances. For the LTM model, we use a scaling relation to find an upper limit on the total cluster mass of M$_{200} \sim 10^{15}$ M$_{\odot}$. We fit a NFW to the WSLAP+ mass profile within 15\arcsec to find a mass of  M$_{200} \approx 6 \times 10^{14}$ M$_{\odot}$.

In summary, CLIO is an excellent cluster for future follow up work. It is a highly concentrated massive system that already has, with modest MUSE spectroscopy, several high-z lensing candidates. Because of its high concentration it is also in a unique regime for massive clusters. More detailed study will reveal properties of its dark matter halo, including measuring its profile in detail and quantifying its sub-halos. Because of its high concentration and low ICL it is an ideal target for JWST observations and will be part of our GTO effort to study the early universe as part of the Webb Median Deep Fields project starting in 2019.


\section*{Acknowledgements}

The authors acknowledge the anonymous referee for the useful comments that helped in improving the paper.

AG and CC acknowledge funding from the STFC. 
This work was funded in part by NASA JWST Interdisciplinary Scientist grants NAG5-12460 and NNX14AN10G from GSFC to RAW.
JMD acknowledges the support of projects AYA2015-64508-P (MINECO/FEDER, UE), AYA2012-39475-C02-01, and the consolider project CSD2010-00064 funded by the Ministerio de Economia y Competitividad. 

Based on observations collected at the European Organisation for Astronomical Research in the Southern Hemisphere under ESO programme(s) 096.B-0605(A).
This work is based [in part] on observations made with the Spitzer Space Telescope, which is operated by the Jet Propulsion Laboratory, California Institute of Technology under a contract with NASA.
Based [in part] on data collected at the Subaru Telescope and retrieved from the HSC data archive system, which is operated by Subaru Telescope and Astronomy Data Center at National Astronomical Observatory of Japan.

We would like to thank the UDSz team (ESO Large Programme 180.A-0776) for providing spectral templates.
We extend our gratitude to Samuel R. Hinton for his support and guidance regarding the modification of the \textsc{marz} software.




\bibliographystyle{mnras}
\bibliography{reference} 



\appendix

\section{WSLAP+ Lensing Constraints}
\label{apx:lensing}

\begin{figure}
	\includegraphics[width=\columnwidth]{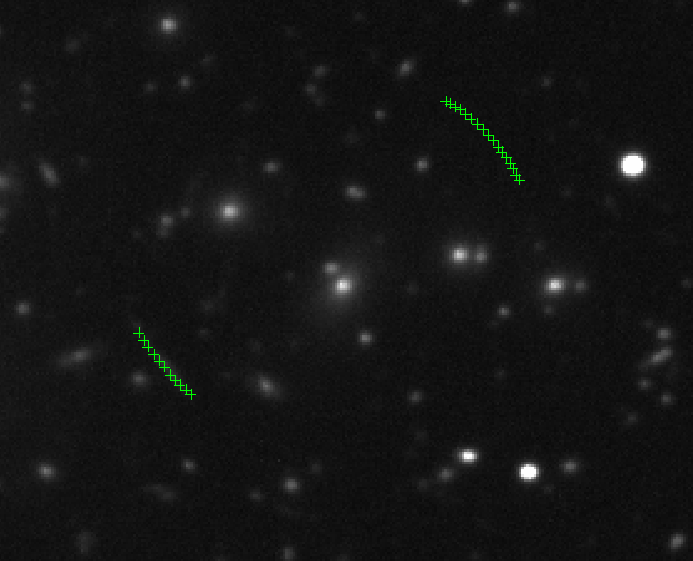}
    \caption{Positions of the A1 (11) and A2 (17) arclet constraints used in WSLAP+ modelling. Points are overlaid on FORS2 r-band image.}
    \label{fig:arcconst}
\end{figure}

For the WSLAP+ lens modeling we employ arclet positions, their extent, and redshifts as constraints to derive a strong lensing mass model. Instead of assuming the centroid positions of the arclets we consider 11, and 17 points for the arcs at $z_s = 2.37$ and $z_s = 2.18$ respectively. We present an illustration of the points used (Figure~\ref{fig:arcconst}), along with a table of central arc positions (Table~\ref{tab:arcconst}). The points are roughly equally spaced with a separation of $\approx 0.7''$ between them. The points do not correspond to particular features in the observed arcs but rather they trace the shape and extension of the arcs.

\begin{table}
  \caption{Table of central positions and redshifts for arclet constraints used in the WSLAP+ modelling.}
  \label{tab:arcconst}
  \begin{tabular}{lcccc}
    \hline
    ARC & id & z & ra (Deg) & dec (deg) \\
    \hline
    A1   & 271 & 2.37 & 130.5960377 & 1.6383164\\
    A2.1 & 153 & 2.18 & 130.5881129 & 1.6458672\\
    A2.2 & 180 & 2.18 & 130.5862754 & 1.6438599\\
    \hline
  \end{tabular}
\end{table}


\bsp	
\label{lastpage}
\end{document}